\documentclass[document]{jfm}
\pdfoutput=1
\usepackage{graphicx}
\usepackage{newtxtext}
\usepackage{newtxmath}
\usepackage{natbib}
\usepackage{physics}
\usepackage[normalem]{ulem}
\usepackage[section]{placeins}
\usepackage{hyperref}
\hypersetup{
    colorlinks = true,
    urlcolor   = blue,
    citecolor  = black,
}

\newcommand{\RomanNumeralCaps}[1]

\newcommand{\ensmean}[1]{\langle #1 \rangle}

\newcommand{\enstmean}[1]{\langle \overline{#1} \rangle}

\title{Competing flow and collision effects in a monodispersed liquid-solid fluidized bed at a moderate Archimedes number}

\author{Yinuo Yao\aff{1,2},
 \corresp{\email{yaoyinuo@stanford.edu}}
 Craig S. Criddle\aff{2}
 \and Oliver B. Fringer\aff{1} }

\affiliation{\aff{1}The Bob and Norma Street Environmental Fluid Mechanics Laboratory, Department of Civil and Environmental Engineering, Stanford University, Stanford, CA, 94305, USA\aff{2}Codiga Resource Recovery Center at Stanford, Department of Civil and Environmental Engineering, Stanford University, Stanford, CA, 94305, USA}

\begin{document}
\maketitle

\begin{abstract}
We study the effects of fluid-particle and particle-particle interactions in a three-dimensional monodispersed reactor with unstable fluidization. Simulations were conducted using the Immersed Boundary Method (IBM) for particle Reynolds numbers of 20-70 with an Archimedes number of 23600. Two different flow regimes were identified as a function of the particle Reynolds number. For low particle Reynolds numbers ($20 < Re_p < 40$), the porosity is relatively low and the particle dynamics are dominated by interparticle collisions that produce anisotropic particle velocity fluctuations. The relative importance of hydrodynamic effects increases with increasing particle Reynolds number, leading to a minimized anisotropy in the particle velocity fluctuations at an intermediate particle Reynolds number. For high particle Reynolds numbers ($Re_p > 40$), the particle dynamics are dominated by hydrodynamic effects, leading to decreasing and more anisotropic particle velocity fluctuations. A sharp increase in the anisotropy occurs when the particle Reynolds number increases from 40 to 50, corresponding to a transition from a regime in which collision and hydrodynamic effects are equally important (Regime 1) to a hydrodynamic-dominated regime (Regime 2). The results imply an optimum particle Reynolds number of roughly 40 for the investigated Archimedes number of 23600 at which mixing in the reactor is expected to peak, which is consistent with reactor studies showing peak performance at a similar particle Reynolds number and with a similar Archimedes number. Results also show that maximum effective collisions are attained at intermediate particle Reynolds number. Future work is required to relate optimum particle Reynolds number to Archimedes number.
\end{abstract}


\section{Introduction}
Fluidization has been widely found in many industrial processes such as water and wastewater treatment and chemical synthesis. In these processes, one of the key operating parameters is the flow rate which in turn controls the porosity to achieve efficient mixing and mass transfer. Depending on the processes, the optimal operating parameter can vary significantly. By gaining a detailed understanding of the particle dynamics in the fluidized bed, system optimization would be possible. Using wastewater treatment as an illustrative example, the Staged Anaerobic Fluidized-bed Membrane Bioreactor (SAF-MBR) was recently proposed to reduce energy demand~\citep{Shin2011-ni,Shin2012-kj,Shin2014-ad} in which granular activated carbon (GAC) particles are fluidized to both maximize biological degradation rate through maximizing the growth of biofilm and minimize membrane fouling through particle collisions. Both objectives can be optimized by choosing the optimal flow rate which in turn controls the porosity. In addition to operation, elucidating the particle dynamics in fluidized bed can provide information to model biofilm detachment rate. Important biofilm detachment mechanisms include shear stress from fluid-particle interactions and abrasion from particle-particle interactions~\citep{Rittmann2018-do,Nicolella1996-fb,Nicolella1997-tx}. However, the proposed models either fail to decouple effect of shear stress from abrasion or only consider abrasion effects~\citep{Chang1991-zw, Nicolella1997-tx,Gjaltema1997-eo}. Without a thorough understanding of the particle dynamics, especially the competition between abrasion and hydrodynamic effects in the fluidized bed, optimizing operating parameters for efficient mixing and modeling biofilm detachment remains an elusive problem.

In the past few decades, many researchers have studied the hydrodynamics of fluidized-bed reactors~\citep{Verma2014-fv, Verma2015-ds, Lu2020-ob, Yang2017-ig, Zenit1997-wm, Duru2002-vj, Derksen2007-mw, Di_Felice1999-xh, Shajahan2020-vx}. In general, fluidized beds are classified into two different categories as either aggregative or particulate~\citep{Geldart1973-uw}. The behavior of fluidized beds can be further classified based on the Froude number $Fr_m = u_{mf}^2/(g d_p)$, where $u_{mf}$ is the minimum fluidization velocity, $g$ is the gravitational acceleration and $d_p$ is the particle diameter~\citep{Sundaresan2003-yp}. For $Fr_m \sim \mathcal{O}(10^{-3})$, no bubbles are observed. For $Fr_{m} \sim \mathcal{O}(10^{-2})$, bubbles appear intermittently. For $Fr_{m} \sim \mathcal{O}(10^{-1})$, bubble-like voids persist. In wastewater treatment, $Fr_{m} \sim \mathcal{O}(10^{-3})$ is generally assumed based on the particles used~\citep{Shin2014-ad}.~\citet{Zenit1997-wm} measured the particle pressure (collision stress) for particles with different properties over a range of volume fractions and showed that the particle pressure initially increases with decreasing volume fraction when the volume fraction is large and then decreases with further decreasing volume fraction, an effect that was also demonstrated by~\citet{Derksen2007-mw}. Since particle pressure measures the effect of collisions, it can be used to quantify abrasion that leads to biofilm detachment. Although they did not directly quantify hydrodynamic stresses,~\citet{Derksen2007-mw} attempted to evaluate the importance of particle streaming stress which is a measure of the hydrodynamic pressure induced by particle velocity fluctuations.~\citet{Yao2021-ex} studies the effect of Archimedes number in concentrated suspensions and discovers that low Archimedes number suspensions result in long-lived particle clusters while high Archimedes number suspensions mainly consist of short-lived particle clusters.

To elucidate the effect of porosity on particle dynamics in the AFBR, we study the fluid-particle and particle-particle interactions in a fluidized bed with the Eulerian-Lagrangian (EL) method. EL methods solve for the dynamics of individual particles and compute the Eulerian flow-discrete particle and particle-particle interactions with different collision models~\citep{Yu2003-zk, Akiki2016-ek, Costa2015-ze, Biegert2017-ku}. EL methods can be further classified into two sub-categories: (i) Volume-averaged Computational Fluid Dynamics-Discrete Element Methods (CFD-DEM) and (ii) Particle Resolved Simulations (PRS). The CFD-DEM method employs various closure laws to model the momentum transfer between the discrete particles and the fluid~\citep{Yu2003-zk, Pan2016-wu}. One advantage of CFD-DEM is its ability to simulate millions of particles, as in the study of the macroscopic behavior of particle-laden flows leading to clustering~\citep{Akiki2017-iq, Akiki2017-mm}. However, there is no consensus on the most appropriate closure laws to parameterize the fluid-particle interactions~\citep{Yin2007-eb,Tenneti2011-eh}. Another disadvantage is that the Eulerian fields are volume averaged which precludes statistical analysis of motions related to detailed fluid-particle interactions.

Because of the need for closure laws in CFD-DEM, it cannot be used
to study the detailed physics of fluid-particle and particle-interactions
in an AFBR. Therefore, we employ the PRS approach which can be thought of as the limiting case of the CFD-DEM
method in which the Eulerian flow is computed on a grid in which many
(15-30) grid cells resolve the smallest particle diameter to simulate
all of the spatio-temporal scales related to the flow-particle
interactions~\citep{Esteghamatian2017-hc,Lee2011-bj,Lee2010-mr}. As a
result, the Eulerian quantities are no longer volume-averaged over
many particles, which allows for a direct quantification of particle
dynamics in systems like fluidized-bed reactors. Three
popular approaches used to study particle-laden flows are the
PHYSALIS method~\citep{Zhang2005-tn}, the Immersed Boundary Method (IBM), and
the Lattice-Boltzmann (LB) technique with comparable
accuracy~\citep{Finn2013-tc}. Recently, the
PRS approach with IBM has been widely applied to a number of different problems, including extracting
drag laws from arrays of particles~\citep{Tenneti2011-eh, Akiki2017-mm,Tang2015-zh} and understanding the detailed physics of
flow-particle interactions in fluidized beds and particle suspensions~\citep{Esteghamatian2017-hc, Kriebitzsch2013-qc, Yin2007-eb, Willen2019-rm, Ozel2017-ro, Uhlmann2014-ja, Zaidi2015-go}.

In this paper, we present PRS results of a fluidized bed reactor to gain a detailed understanding of the
effects of varying upflow velocity on the particle dynamics. A series of cases with different
particle Reynolds numbers is studied and the simulation results
are used to (1) understand the
equilibrium behavior of particle fluctuations, (2) establish links
between particle velocity fluctuations, forces on particles and flow-particle microstructure
and (3) identify various flow regimes and momentum transfer
mechanisms as a function of the particle Reynolds number.

\section{Numerical methodology}

\subsection{Equations and discretizations}
The governing Navier-Stokes equations are solved in a three-dimensional
flow reactor with a square cross section and with an array of uniform spherical particles. Direct
forcing by the IBM method is accounted for with a source term,
$\boldsymbol{f}_{IBM}$, which is added to the Navier-Stokes equation
and enforces no-slip boundary conditions on the particle surfaces.
With this forcing, the incompressible Navier-Stokes equation we solve is
given by
\begin{eqnarray}
\pdv{\boldsymbol{u}}{t} + \boldsymbol{u} \vdot \grad{\boldsymbol{u}} = -\grad{p} + \nu \laplacian{\boldsymbol{u}} + \boldsymbol{f}_{IBM}, \label{NS-eqs} 
\end{eqnarray}
subject to continuity, $\div{\boldsymbol{u}} = 0$, where $\boldsymbol{u}$ is the
velocity vector and $p$ is the pressure normalized by the fluid
density, $\rho_f$. These equations are discretized on a uniform collocated Cartesian grid and momentum and pressure are coupled with the fractional step method~\citep{Zang1994-ck}. The advection term is discretized with the
explicit, three-step Runge-Kutta schemes described in~\citet{Rai1991-dp}. The viscous term is discretized with the
implicit Crank-Nicolson scheme to eliminate the associated stability
constraints. 

The IBM formulation employs the direct forcing approach first proposed
by~\citet{Uhlmann2005-hf} and improved by~\citet{Kempe2012-lp}. This approach represents the particle
using $N_L$ Lagrangian markers with volume $\Delta{V}_L\approx\Delta{V}_E$,
where $\Delta{V}_L$ is the Lagrangian marker volume and $\Delta{V}_E$
is the volume of each Eulerian grid cell. At each time step, the IBM force,
$\boldsymbol{f}_{IBM}$, is calculated as the force required to enforce the
difference between the desired velocity ${\bf u}_d$ at the particle surface
and the interpolated velocity from the Eulerian grid. The desired velocity
at the particle surface is calculated as the sum of a translational (${\bf u}_p$) and a rotational
components based on the particle angular velocity vector ${\bf \omega}_p$ 
with ${\bf u}_d = {\bf u}_p + {\bf\omega}_p\times{\bf r}$, where ${\bf r}$ is the
vector pointing from the particle center of mass to the Lagrangian point.
The translational and angular velocities of the particle are then governed by
\begin{subequations}
\label{par-eqs} 
\begin{eqnarray}
m_p \dv{\boldsymbol{u}_p}{t} &=& \rho_f\Big[\dv{}{t}\int_{\Omega_p}  \boldsymbol{u} \ \dd \Omega_p - \int_{S} \boldsymbol{f}_{IBM} \ \dd S \Big] + V_p(\rho_p - \rho_f)\boldsymbol{g} + \boldsymbol{F}_{c,p},  \label{eq:up}\\
I_p \dv{\boldsymbol{\mathbf{\omega}}_p}{t} &=& \rho_f\Big[\dv{}{t}\int_{\Omega_p} \boldsymbol{r} \times \boldsymbol{u} \ \dd \Omega_p - \int_{S} \boldsymbol{r} \times \boldsymbol{f}_{IBM} \ \dd S \Big] + \boldsymbol{T}_{c,p},\label{eq:omegap}
\end{eqnarray}
\end{subequations}
where the integrals are calculated over the discrete volumes associated with the Lagrangian points
given by the surface $S$ and region $\Omega_p$, $m_p$ is the mass of the particle,
$\boldsymbol{g}$ is the gravitational acceleration vector which points in the negative z-direction,
$V_p$ is the volume of the particle, $\rho_p$ is the particle density,
$\boldsymbol{F}_{c,p}$ and $\boldsymbol{T}_{c,p}$ are the force and torque due to
particle-particle and particle-wall interactions, and $I_p$ is the moment of inertia of the
particle. For a description of each term in equations~(\ref{par-eqs}), please refer to the papers by~\citet{Uhlmann2005-hf} and~\citet{Kempe2012-lp}. Interpolation of variables between the
Lagrangian and Eulerian grid employs the discrete delta function
kernel proposed by~\citet{Peskin1977-wj} and~\citet{Roma1999-cp}.

The original direct forcing approach proposed by~\citet{Uhlmann2005-hf} has two
disadvantages: (1) the rigid-body approximation used to approximate the volume integral, $\dv{}{t}\int_{\Omega_p} \rho_f \boldsymbol{u} \ \dd \Omega_p$ in equation~(\ref{par-eqs}) introduces a singularity at density ratio of $\rho_p/\rho_f=1$ and stability limit of
$\rho_p/\rho_f \approx 1.2$~\citep{Uhlmann2005-hf,Kempe2012-lp} and (2) the explicit calculation of $\boldsymbol{f}_{IBM}$ leads to error that is inversely proportional to $Re_p$ and cannot be eliminated by refining the time step~\citep{Kempe2012-lp}.
The restriction associated with the rigid-body approximation
is eliminated by introducing the numerical level-set approximation of the volume integral in equation~(\ref{par-eqs}) instead of applying the rigid-body approximation.
The error associated with $\boldsymbol{f}_{IBM}$ is reduced by applying a heuristic
number of outer forcing loops~\citep{Kempe2012-lp,Wang2008-fx}. In each
outer forcing loop, the Eulerian velocity field is interpolated
between the Eulerian and Lagrangian grid and updated with the newly
calculated $\boldsymbol{f}_{IBM}$ to satisfy the no-slip boundary condition at
the particle surface. Details can be found in the paper by~\citet{Kempe2012-lp}, who recommended three
outer forcing loops while~\citet{Biegert2017-ku}
recommended one loop based on various validation cases. After validating with different
cases, two outer forcing loops were used for the simulations in this study.

To model the forces associated with particle-particle and particle-wall interactions, the total force and torque on particle $p$ due to interaction with $q=1,2,\dots,N_p$ particles and a wall $w$ are computed with
\begin{subequations}
\label{f_int}
\begin{eqnarray}
\boldsymbol{F}_{c,p} &=&  \sum_{p, q \ne p}^{N_p} (\boldsymbol{F}_{n,q} + \boldsymbol{F}_{t,q}) + \boldsymbol{F}_{n,w} + \boldsymbol{F}_{t,w}, \\
\boldsymbol{T}_{c,p} &=&  \sum_{p, q \ne p}^{N_p} R_{p,cp} \boldsymbol{n}_{p,q} \times \boldsymbol{F}_{t,q} + R_{p,cp} \boldsymbol{n}_{p,w} \times \boldsymbol{F}_{t,w},
\end{eqnarray}
\end{subequations}
where $\boldsymbol{F}_{n,q}$ and $\boldsymbol{F}_{t,q}$ are the normal and tangential
collision forces between particle $p$ and $q$, $\boldsymbol{F}_{n,w}$ and
$\boldsymbol{F}_{t,w}$ are the normal and tangential collision forces between
particle $p$ and a wall, $R_{p,cp}$ is the effective radius between particle $p$
and $q$, $\boldsymbol{n}_{p,q}$ is the vector normal to the plane of contact between particles $p$ and $q$,
and $\boldsymbol{n}_{p,w}$ is the vector normal to the wall at the point of contact with particle $q$. Normal collision forces $\boldsymbol{F}_{n,q}$ consist of contributions from both lubrication and contact based on the separation distance between the particles~\citep{Biegert2017-ku}. Collision models usually have high stiffness in which the collision time step size is much smaller than the fluid time step size. To avoid this, we follow the approach by~\citet{Biegert2017-ku} who employ
the collision model proposed by~\citet{Kempe2012-pl}.

The code we employ for the simulations in this paper is based on the code developed by Drs. Hyungoo Lee and Sivaramakrishnan Balachandar, who employed IBM with the direct forcing approach by~\citet{Uhlmann2005-hf} to simulate the near-wall motion of an isolated particle~\citep{Lee2010-mr,Lee2011-bj}. We modified their code following improvements to the IBM method suggested by~\citet{Kempe2012-lp} and~\citet{Biegert2017-ku} as follows. To improve the accuracy of computing fluid-solid interactions, the following modifications were made: 1) Three-step Runge-Kutta time-stepping scheme instead of third-order Adams-Bashforth for the advection terms in equation~(\ref{NS-eqs}), 2) Outer forcing loops to reduce the error associated with explicit calculation of the forcing in the IBM method~\citep{Kempe2012-lp} and 3) Higher-order schemes with predictor and corrector steps to discretize the particle motion equations (equation~\ref{par-eqs})~\citep{Biegert2017-ku}. Because the original code was not designed to simulate particle-particle interactions, collision models based on~\citet{Kempe2012-pl} and~\citet{Biegert2017-ku} were implemented. Finally, computational performance was made efficient by 1) using appropriate MPI (Open MPI-3.0) structures to transfer Lagrangian and Eulerian information related to particle-particle collisions at interprocessor boundaries 2) employing Hypre libraries developed at the Lawrence Livermore National Laboratories~\citep{Falgout2006-bt,Chow1998-me} to solve the linear systems associated with implicit discretization of the viscous terms and the pressure-Poisson equation. 

\section{Fluidized bed simulation setup}
\begin{figure}
\centering
\includegraphics[height = 0.5\textheight, keepaspectratio]{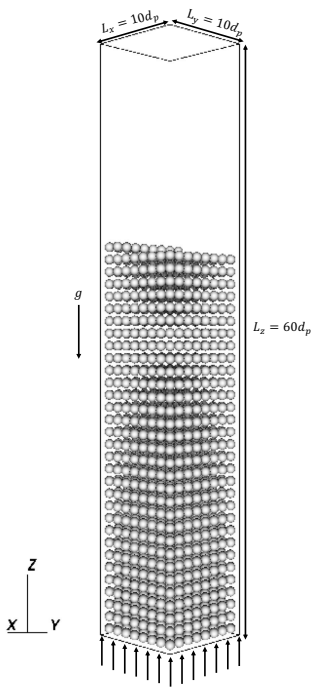}%
\caption{\label{sch_diag} The three-dimensional
 computational domain, showing the fluidized bed, the uniform inflow velocity profile and the initial particle positions.}
\end{figure}

To simulate the fluidized bed, three-dimensional simulations are conducted with $N_p=2000$ particles in the
reactor channel shown in figure~\ref{sch_diag}. The particles have a uniform diameter $d_p=2$~mm and density $\rho_p=1300$~kg~m$^{-3}$,
and the fluid has a kinematic viscosity $\nu=10^{-6}$~m$^2$~s$^{-1}$
and density $\rho_f=998.21$~kg~m$^{-3}$, resulting in Archimedes number $Ar=23600$ (Galileo number $Ga = 154$) that is defined as
\begin{eqnarray}
\label{eq:Ar-def}
&& Ar = Ga^2 = \dfrac{g(s-1)d_p^3}{\nu^2},
\end{eqnarray}
where $s = \rho_p/\rho_f$ is the ratio of particle to fluid density. The particle collisions have a dry restitution coefficient $e_{dry}=0.97$~\citep{Joseph2001-wa,Foerster1994-ct}, coefficient of kinetic friction $\mu_k=0.15$~\citep{Joseph2004-pq} and coefficient of static friction $\mu_s=0.8$~\citep{Dieterich1972-kx}. The grid spacing is uniform
in the $x$, $y$ and $z$ directions and given by $\Delta x= \Delta y = \Delta z=h=d_p/25.6$,
which is sufficient to resolve the flow-particle interactions~\citep{Biegert2018-rh,Kempe2012-lp}. The channel width is given by $L_x = L_y = 10 d_p$ and its length is $L_z = 60 d_p$, giving
a three-dimensional grid with 256$\times$256$\times$1536 grid points.
The time-step size is $\Delta{t}=1.5\times10^{-4}$~s, resulting in a maximum Courant number of
$0.5$ for the cases with the highest upflow velocities.

The primary parameter of interest is the particle Reynolds number $Re_p = u_0 d_p/\nu$,
where the average upflow velocity at the inlet, $u_0$, is varied to investigate Reynolds number
effects. Cases are run with periodicity in the $x$ and $y$ directions. A total of six simulations were conducted with
$0.010\le u_0\le 0.035$~m~s$^{-1}$, giving $20 \le Re_p \le 70$. For all cases, the pressure is specified at the top boundary as $p=0$,
while at the bottom boundary the inflow velocity is specified as uniform and given by $w(x,y)=u_0$.

Simulations are initialized with a uniform distribution
of particles with a spacing of $1 d_p$ and the flow is
impulsively started from rest. The upflow velocity leads to expansion
of the bed and random motion of the particles until statistical
equilibrium is reached, at which time the dynamics are independent of
the initial particle distribution. 

\section{Results and discussion}

\subsection{Instantaneous and time-average porosity} \label{sec:steady_stats} 
\begin{figure}
\centerline{\includegraphics[width=\textwidth]{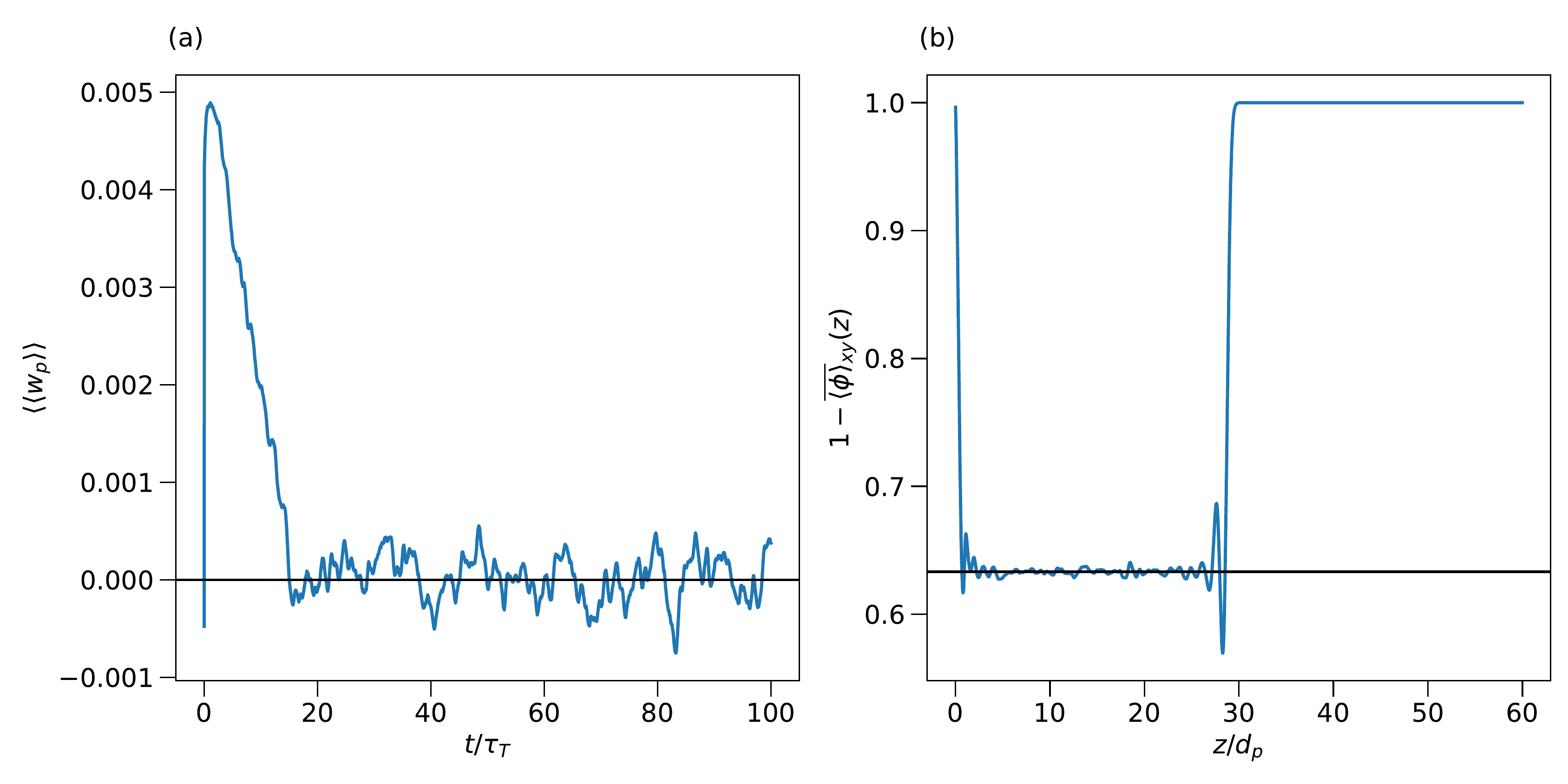}}
\caption{\label{poro_t}
 (a) Time series of the ensemble-average instantaneous vertical particle velocity $\langle \langle w_p \rangle \rangle$ showing how statistical equilibrium is reached
 at roughly $t=20\tau_T$, where $\tau_T=d_p/u_0$ is the particle turnover time and (b) vertical variation of time-average porosity $1 - \enstmean{\phi}_{xy}$ showing the effects of boundary conditions on the bottom and top of the fluidized bed for the case with $Re_p = 40$.}
\end{figure}

To understand the time evolution of the particles, we define the ensemble-average instantaneous vertical velocity of particles using the ensemble-average
\begin{eqnarray}
    \langle \langle \{\cdot\} \rangle \rangle = \dfrac{1}{N_p} \sum_{i=1}^{N_p} \{\cdot\}_i,
\end{eqnarray}
where $\{\cdot\}_i$ corresponds to the variable of particle $i$. As shown in figure~\ref{poro_t}(a), the ensemble-averaged instantaneous vertical velocity of particles $\langle\langle w_{p} \rangle\rangle$ initially
increases with time as the bed expands because the average drag force
on the particles exceeds their submerged weight. Eventually, the
average drag is in balance with the submerged weight, leading to
statistical equilibrium. Defining the particle turnover time as $\tau_T=d_p/u_0$, simulations are run for $t_{max}=100-300\tau_T$, depending on $Re_p$, to ensure statistically-stationary
results. Time-averaged statistics are denoted by the overbar and computed over the last 80 turnover times,
such that
\begin{eqnarray}
\label{time_average}
 \overline{\{ \cdot \}}=\frac{1}{80\tau_T}\int_{t_0}^{t_{\max}}\{ \cdot \}\,\dd t\,,
\end{eqnarray}
where $t_0=t_{\max}-80\tau_T$. Unless otherwise stated, the time-averaging operator has been applied to compute statistically-stationary quantities.

In the simulations, the porosity decreases from unity at the inlet, remains relatively constant and then increases to unity at the top of the fluidized bed. Therefore, a spatially variable porosity is expected as compared to the homogeneous porosity in a fluidized bed away from boundaries. Including results from these two regions (top and bottom) would affect the accuracy and convergence of statistical quantities. To determine the region of homogeneous porosity, we define the instantaneous Eulerian volume fraction of the bed, $\phi(\boldsymbol{x},t)$ and compute horizontally-averaged (x- and y-directions) Eulerian volume fraction of the bed, $\enstmean{\phi}_{xy}(z,t)$ as described in Appendix~\ref{sec:appA}. Figure~\ref{poro_t}(b) shows the time-averaged porosity $1-\enstmean{\phi}_{xy}$ as a function of the vertical position $z/d_p$ for $Re_p=40$. The time-averaged porosity decreases to $0.63$ away from the inlet and then increases to $1$ at the top of the fluidized bed. To eliminate boundary effects, we define the modified spatial average for fluid variables as
\begin{eqnarray}
    \label{mod_spa_average}
  &&  \langle\{ \cdot \} \rangle_\beta = \frac{1}{N_\beta} \sum_{i=1}^{N_\beta}  \{ \cdot \}_{ijk}, \\
  && \langle\{ \cdot \} \rangle_z = \frac{1}{N_z^*} \sum_{k=k_s}^{k_e}  \{ \cdot \}_{ijk}, \\
   && \langle\{ \cdot \} \rangle = \frac{1}{N_xN_yN_z^*} \sum_{i,j=1}^{N_x,N_y} \sum_{k=k_s}^{k_e}  \{ \cdot \}_{ijk},
\end{eqnarray}
where $\beta = x$ or $y$ represents the horizontal directions, $k_s = z_{b}/h$ and $k_e = z_{t}/h$ are the nearest integer of the bottom and top of the fluidized-bed and $z_b$ and $z_t$ are the vertical position of the bottom and top of the homogeneous fluidized bed, respectively. The modified ensembled-average operator for particle variables is then defined as 
\begin{eqnarray}
    \label{par_mod_ens_average}
    \langle\{ \cdot \} \rangle = \frac{1}{N_p^*} \sum_{i=1}^{N_p} \{ \cdot \}_i\mathbf{1}_{z_{b} < z_p < z_{t}}(z_p),
\end{eqnarray}
where $N_p^* = \sum_{n=1}^{N_p} \mathbf{1}_{z_{b} < z_p < z_{t}}(z_p)$ is the number of particles that are located within the homogeneous fluidized bed and  
\begin{eqnarray}
\mathbf{1}_{z_{b} < z_p < z_{t}}(z_p) = 
\begin{cases}
1 & z_b < z_p < z_t, \\
0 & \text{otherwise,}
\end{cases}
\end{eqnarray}
is the indicator function that describes whether particles are located in the spatially homogeneous region of the  fluidized bed.

\begin{figure}
\centering
\includegraphics[width=.45\textwidth]{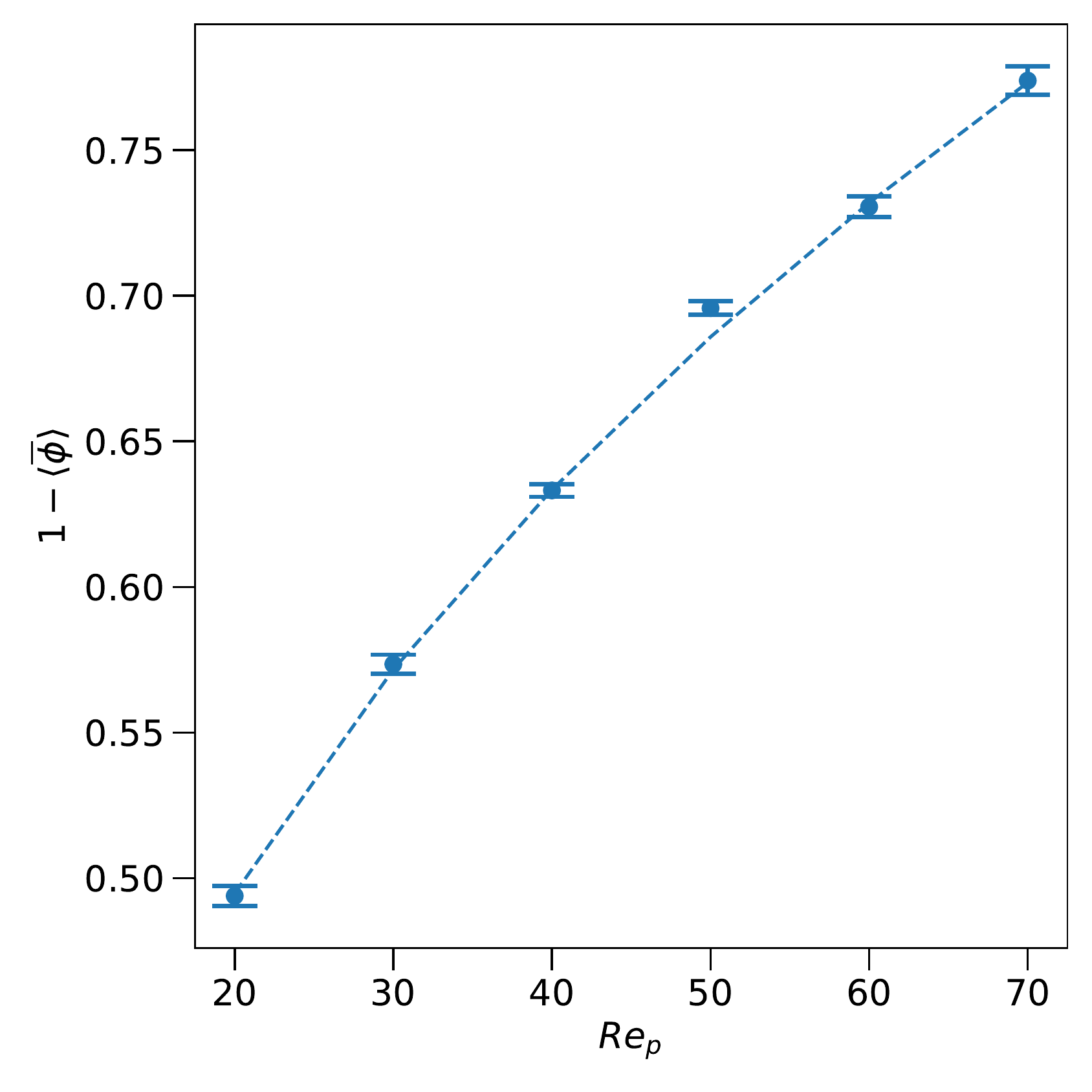}
\caption{\label{expan_index}
Porosity $1-\enstmean{\phi}$ as a function of $Re_p$ for the simulated cases. The dashed lines were constructed based on fitting to the power law equation~(\ref{pow_law}).}
\end{figure}

The relationship between the upflow velocity and volume
fraction of particles has been studied extensively~\citep{Yin2007-eb,Richardson1954-ay, Garside1977-vp, Di_Felice1995-mg, Di_Felice1999-xh, Di_Felice1996-vn, Willen2019-rm, Hamid2014-bi,Zaidi2015-go,Nicolai1995-wi}  and is typically described by the power law relationship
\begin{eqnarray}
u^* = \frac{u_0}{w_{ref}} = k(1-\enstmean{\phi})^n\,, \label{pow_law}
\end{eqnarray}
where $w_{ref}$ is the settling velocity of a single particle in
the domain of interest, $k$ is a low volume fraction correction~\citep{Yin2007-eb,Di_Felice1995-mg, Di_Felice1999-xh, Di_Felice1996-vn} and $n$ is the expansion or power law exponent.The settling velocity of a single particle in an infinitely large domain, $w_{ref}$, is computed with~\citep{Yin2007-eb}
\begin{eqnarray}
\label{eq:ar-ret}
&& Ar = 
\begin{cases}
18 Re_t [ 1 + 0.1315 Re_t^{(0.82-0.05\log_{10}Re_t)}], & 0.01 < \Re_t < 20 \\
18 Re_t [ 1 + 0.1935 Re_t^{0.6305}], & 20 < Re_t < 260 
\end{cases}
\end{eqnarray}
where $Re_t = w_{ref} d_p/ \nu$ and $Ar=(\rho_p/\rho_f -1)gd_p^3/\nu^2$ is the Archimedes number. In this work, $Ar = 2.36\times10^{4}$ giving $Re_t \approx 200$ which is within the range of equation~\ref{eq:ar-ret}. The error of equation~\ref{eq:ar-ret} has been shown to range from 2\% to 4\%~\citep{Yin2007-eb,Willen2019-rm}. Fitting our results to equation~\ref{pow_law} yields values of $k=0.72$ and $n=2.82$ that are consistent with published values~\citep{Yin2007-eb,Willen2019-rm}. In this paper we focus on the relationship between the porosity $1-\enstmean{\phi}$ and $Re_p$ shown in figure~\ref{expan_index}, which also includes the power law fits and shows that the porosity increases with increasing $Re_p$.

\subsection{Kinematic wave speed}
As discussed by many researchers~\citep{Sundaresan2003-yp, Duru2002-vj, Derksen2007-mw,Ham1988-wp}, as $Re_p$ increases above $Re_{mf} = u_{mf}d_p/\nu$ where $Re_{mf}$ is the minimum fluidization Reynolds number, particles start to expand and upward-propagating waves can be observed. Below an intermediate $Re_p$, stable and neutral wave modes exist in which the amplitude of the waves decreases and remains approximately constant. For larger $Re_p$, unstable wave modes develop in which the wave amplitude grows with height, leading to large fluctuations in porosity.~\citet{Zenit2000-cj} investigated the time evolution of the ensemble- and time-averaged porosity $1-\enstmean{\phi}$ and concluded that large-amplitude,  low-frequency fluctuations dominated at low porosity (high volume fraction $\phi > 0.3$) and small-amplitude, high-frequency fluctuations dominated at high porosity (low volume fraction $\phi < 0.3$). 
\begin{figure}
\centerline{\includegraphics[width=\textwidth]{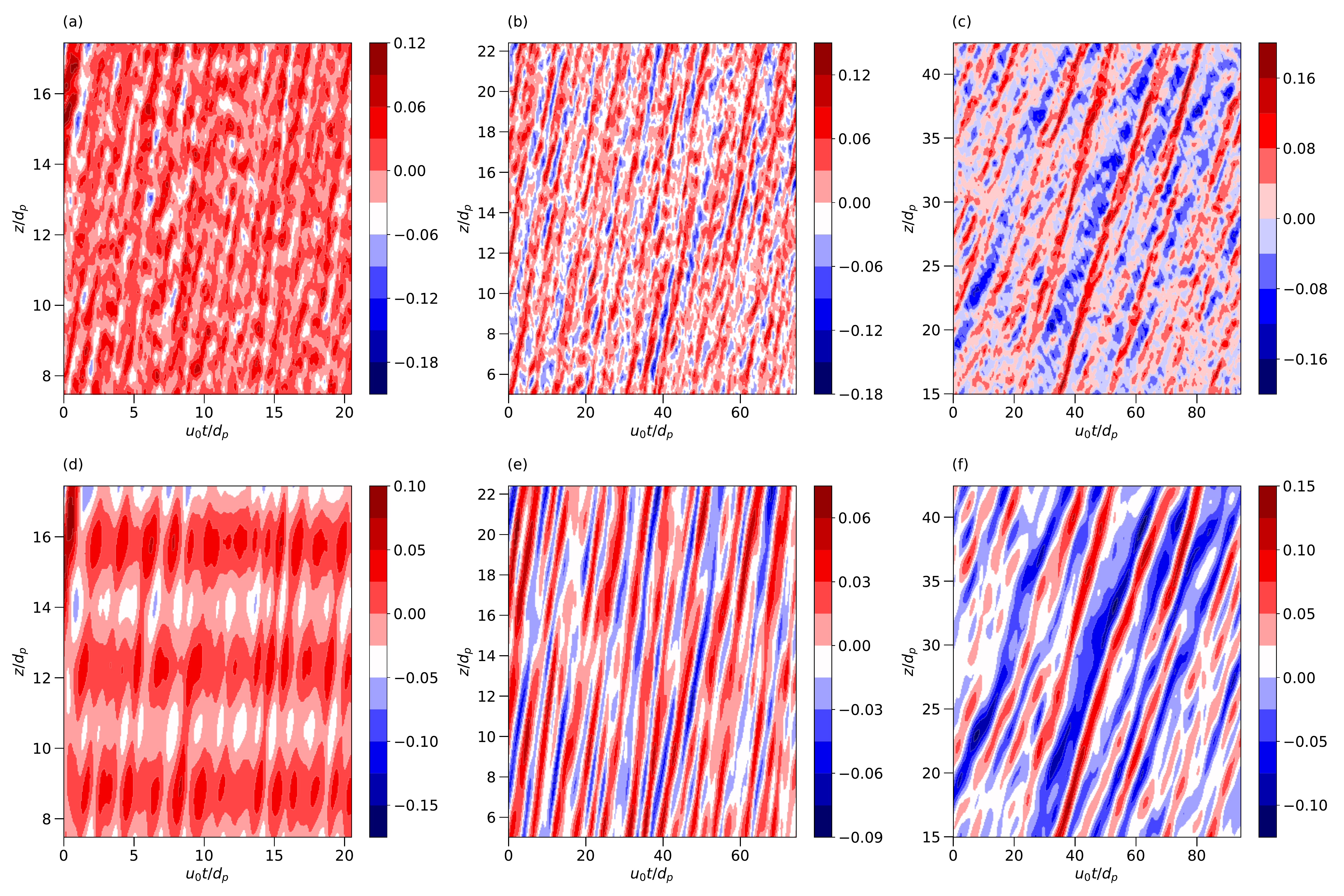}}%
\caption{\label{fig:phi_over_z}Volume fraction fluctuation $\ensmean{\phi}_{xy}^\prime$ as a function of time $u_0t/d_p$ and vertical position $z/d_p$ for the cases with $Re_p = 20$, $40$ and $70$ from left to right. The second row represents the low wavenumber motions with wavenumber $k_z < k_{z,thresh}$.}
\end{figure}

Figure~\ref{fig:phi_over_z}(a), (b) and (c) show two-dimensional $z-t$ plots of the volume fraction fluctuation $\ensmean{\phi}_{xy}^\prime = \ensmean{\phi}_{xy} - \enstmean{\phi}_{xy}$. Waves can be identified visually but are difficult to quantify due to the presence of noise and superposition of waves with different modes. Researchers have reported that $\ensmean{\phi}_{xy}^\prime$ can be classified based on frequencies~\citep{Zenit2000-cj} or wavenumber~\citep{Willen2017-ht}. To separate low-wavenumber waves from the raw data, we modified the Fourier reconstruction method proposed by~\citet{Willen2017-ht} by defining low-wavenumber fluctuations for $k_z < k_{z,thresh}$, where $k_{z,thresh}= (z_t - z_b)/2d_p$ is the threshold wavenumber. Since the behavior of small spatial-scale fluctuations ($k_z \ge k_{z,thresh}$) is an artifact of the spatial averaging, therefore, this study only focuses on the large-scale fluctuations ($k_z < k_{z,thresh}$) which allow interpretation of kinematic wave behavior. Figure~\ref{fig:phi_over_z}(d), (e) and (f) show the low-wavenumber $\phi^\prime_{k_z < k_{z,thresh}}$. In general, upward-propagating waves are clearly observed with alternating regions of high and low porosity for $k_z < k_{z,thresh}$. For $Re_p =20$ (figure~\ref{fig:phi_over_z}(d)), $\phi^\prime_{k_z < k_{z,thresh}}$ is a strong function of vertical position $z$ in which $\phi^\prime_{k_z < k_{z,thresh}}$ varies over a distance of $\sim 2d_p$ and depends weakly on time. For $Re_p > 20$, wave motion is dependent both on time and vertical position. 

\begin{figure}
\centerline{\includegraphics[width=\textwidth]{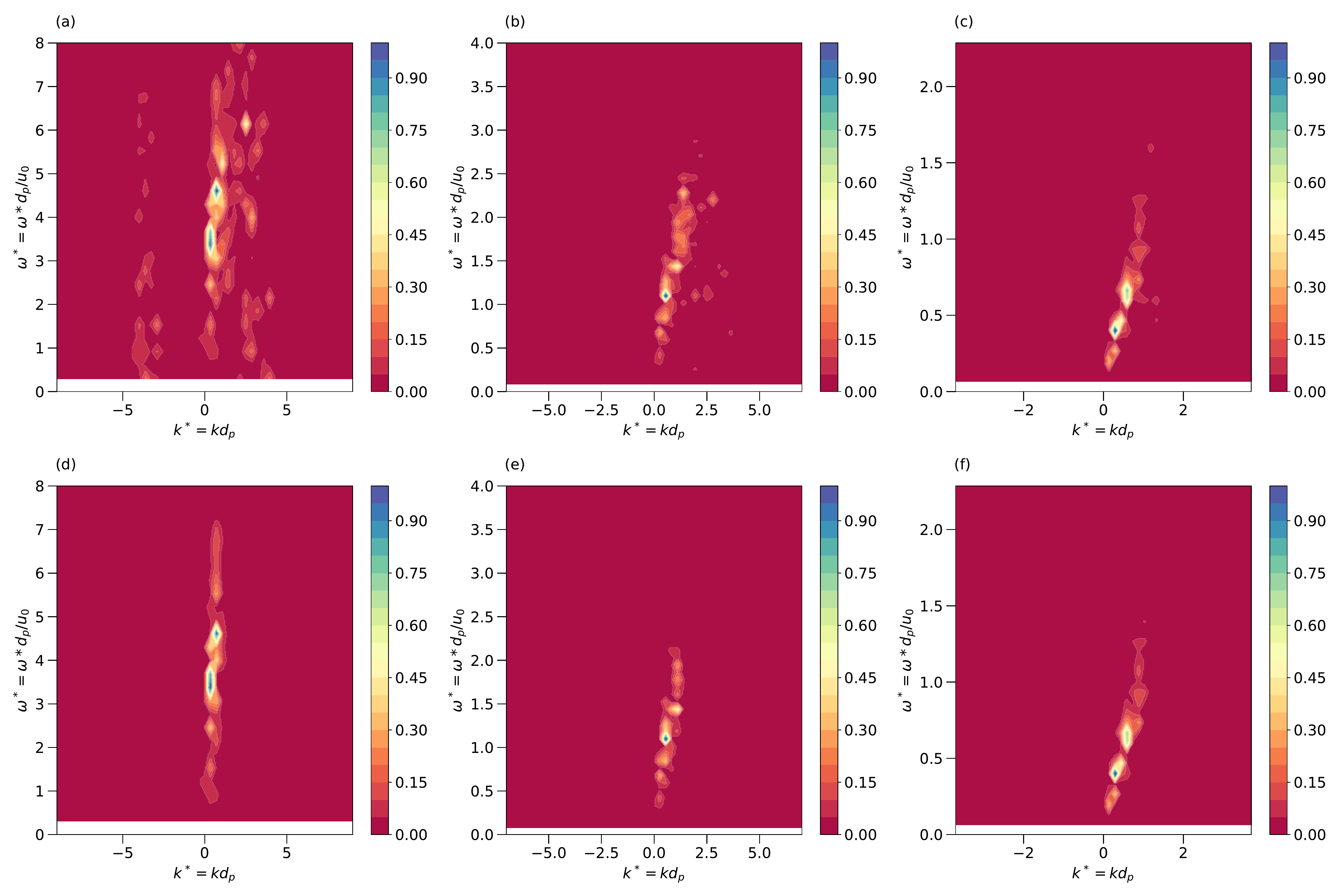}}%
\caption{\label{fig:phi_over_kw}Energy spectra of volume fraction fluctuation $\ensmean{\phi}_{xy}^\prime$ as a function of normalized wavenumber $k^*$ and normalized frequency $\omega^*$ for the cases with $Re_p = 20$, $40$ and $70$ from left to right and with $\phi^\prime$ and  $\phi^\prime_{k_z < k_{z,thresh}}$ from top to bottom.}
\end{figure}

The wave-like motions can be represented by a superposition of different waves with normalized wavenumber $k^*=kd_p$ and normalized frequency $\omega^*=\omega d_p/u_0$. Figure~\ref{fig:phi_over_kw}(a),(b) and (c) show the $\ensmean{\phi}_{xy}^\prime$ in spectral space as a function of $k^*$ and $\omega^*$. Overall, $\ensmean{\phi}_{xy}^\prime$ is dominated by low wavenumber waves with $k^* \approx 1$ and the dominant frequency $\omega^*$ decreases as $Re_p$ increases. At $Re_p = 20$, regions of high $k^*-\omega^*$ can be observed even though the spectral density is relatively small. This illustrates that high-frequency wave modes are more significant at low $Re_p$ than high $Re_p$. As $Re_p$ increases, different modes collapse into a linear relationship. Figure~\ref{fig:phi_over_kw}(d),(e) and (f) show the $k^*-\omega^*$ spectra of $\phi^\prime_{k_z < k_{z,thresh}}$. For $\phi^\prime_{k_z < k_{z,thresh}}$, by eliminating high wavenumber modes, different modes of waves collapse into a line and a wave speed can be estimated from the kinematic relationship $\omega = ck$ using linear regression.

\begin{figure}
\centerline{\includegraphics[width=.5\textwidth]{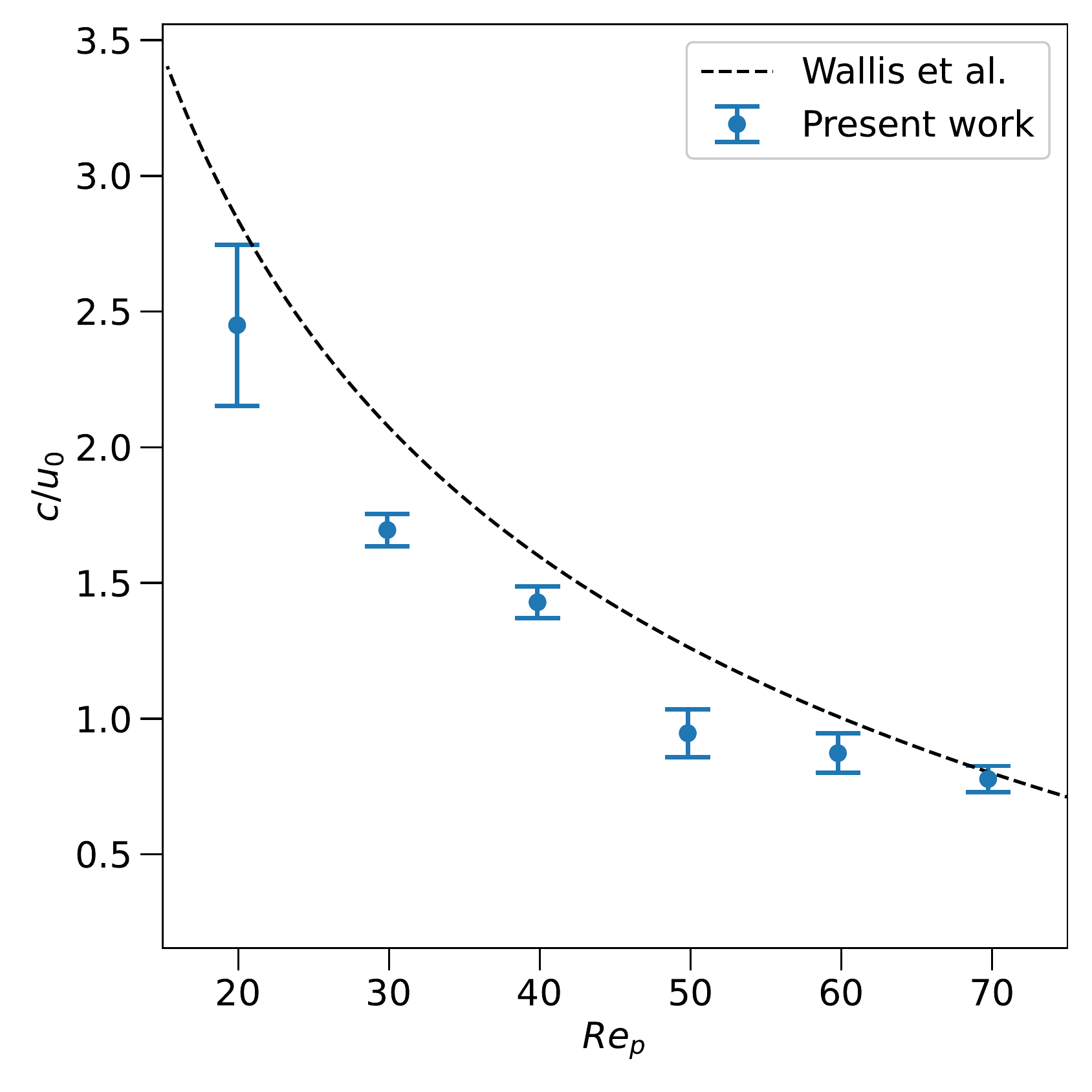}}%
\caption{\label{fig:wavespeed}Wave speed based on the autocorrelation as a function of the particle Reynolds number $Re_p$.}
\end{figure}

To demonstrate the existence of kinematic waves in the fluidized bed, we employ the model proposed by~\citet{Wallis2020-mr} which relates volume fraction to wave speed with 
\begin{eqnarray}
\label{eq:wavemodel}
c = k n \enstmean{\phi} (1-\enstmean{\phi})^{n-1} w_{ref},
\end{eqnarray}
where $c$ is wave speed and other variables are defined in equation~\ref{pow_law}. To calculate the wave speed, we employ the two-dimensional autocorrelation approach as demonstrated by Yao et al. (2021). The autocorrelation of reconstructed volume fraction fluctuation $\phi^\prime_{k_z < k_{z,thresh}}$ is computed and wave speeds are calculated as the slope. As shown in figure~\ref{fig:wavespeed}, the calculated wave speeds are in agreement with the modeled values. The difference between the calculated wave speeds and modeled values are likely due to the dispersive effects which are not considered in equation~\ref{eq:wavemodel}~\citep{Shajahan2020-vx,Wallis2020-mr}.

\subsection{Velocity fluctuations}
\begin{figure}
\centerline{\includegraphics[width=.5\textwidth]{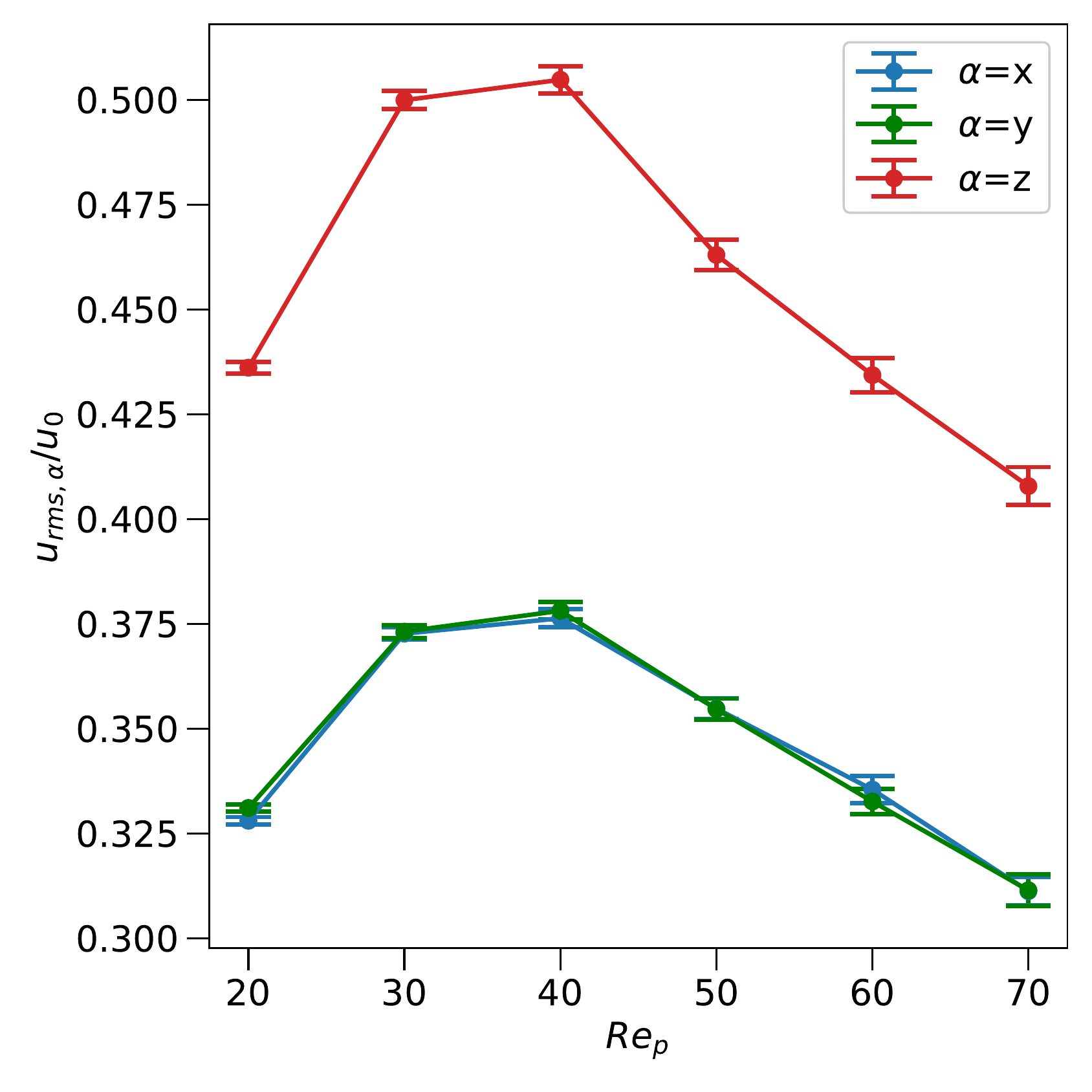}}%
\caption{\label{fig:urms_Rep}The root-mean-square velocity fluctuation normalized by the superficial velocity $u_0$ as a function of the particle Reynolds number $Re_p$.}
\end{figure}

In a fluidized bed, the ensemble-averaged vertical velocity $\langle \langle w \rangle \rangle = 0$, indicating a balance between the submerged weight of the particles and the drag force. However, many researchers have reported that the velocity fluctuations can be as high as $10\%-170$\% of the superficial velocity $u_0$ depending on the volume fractions~\citep{Willen2019-rm,Hamid2014-bi,Zaidi2015-go,Nicolai1995-wi}. To understand the effect of $Re_p$ on the velocity fluctuations, we computed the root-mean-square velocity fluctuation 
\begin{eqnarray}
u_{rms,\alpha} = \overline{\sqrt{\ensmean{u^\prime_\alpha u^\prime_\alpha}}},
\end{eqnarray}
where $u^\prime_\alpha = u_\alpha - \enstmean{u}_\alpha$ is the particle velocity fluctuation and $\alpha = x$, $y$ or $z$. Here, the overbar is the time average defined in equation~(\ref{time_average}) and the angular brackets $\ensmean{\cdot}$ imply an ensemble average over particles in the homogeneous region of the fluidized bed as defined in equation~(\ref{par_mod_ens_average}). Figure~\ref{fig:urms_Rep} shows $u_{rms,\alpha}$ normalized by $u_0$ as a function of $Re_p$. Both $u_{rms,x}$ and $u_{rms,z}$ increase initially and reach a maximum at $Re_p \approx 40$ and then decrease with increasing $Re_p$. Defining the anisotropy as $u_{rms,z}/u_{rms,x}$, for the range of $Re_p$ simulated, the anisotropy ranges between $1.7 - 1.8$. Similar trends were observed by~\citet{Willen2019-rm}. While a maximum in velocity fluctuations is expected because they should be zero for both a single particle ($\phi \approx 0$) and a packed bed ($\phi \approx 0.6$), the physical mechanisms leading to a maximum at $Re_p \approx 40$ have not been reported in the literature. 

\begin{figure}
\centerline{\includegraphics[width=.5\textwidth]{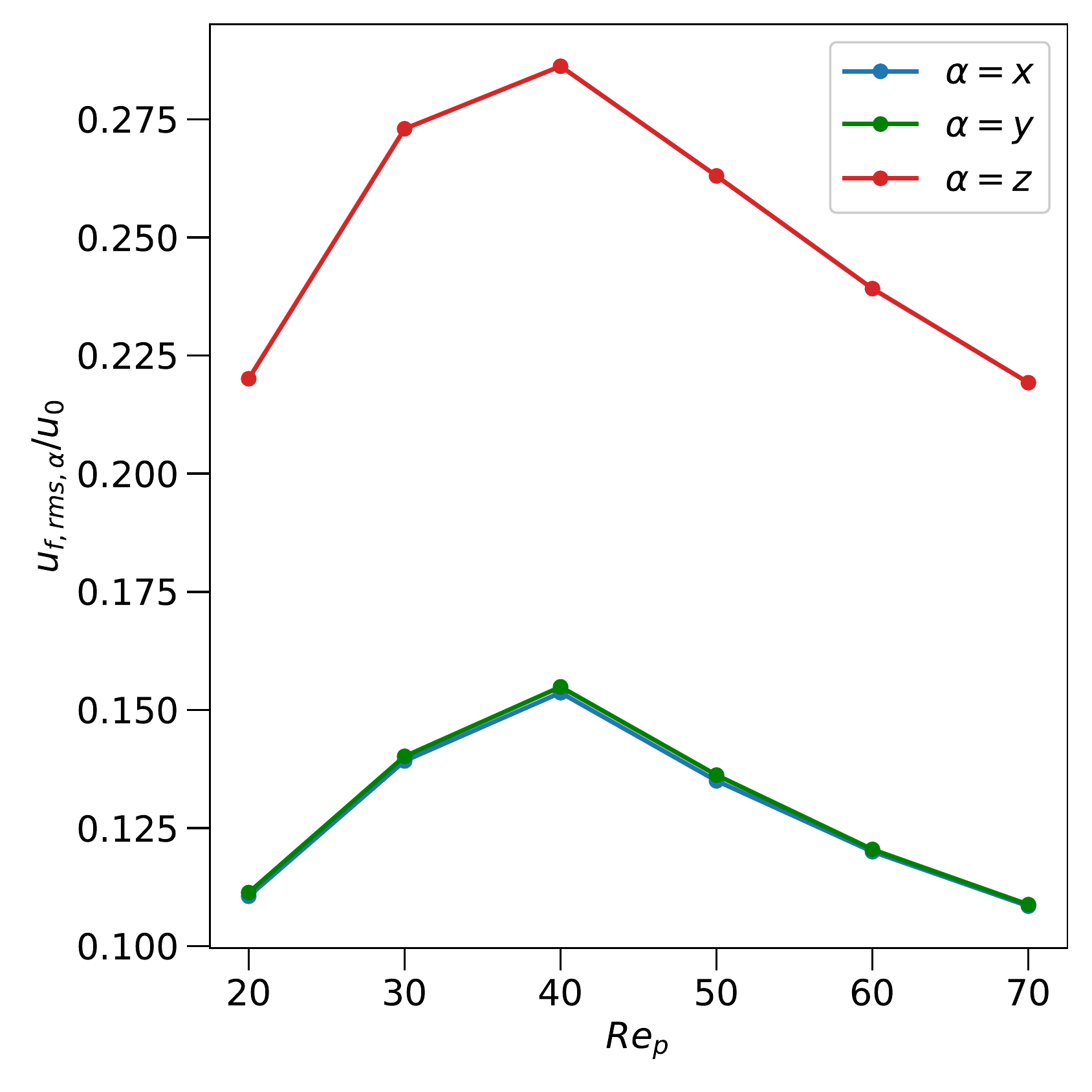}}%
\caption{\label{fig:urms_Rep_fluid}The root-mean-square fluid velocity fluctuation in vicinity to particles normalized by the superficial velocity $u_0$ as a function of the particle Reynolds number $Re_p$.}
\end{figure}

In addition to particle velocity fluctuations, the fluid velocity fluctuations in the vicinity of the particles are also quantified. Many methods have been proposed to quantify the fluid velocity in the vicinity of the particles~\citep{Bagchi2003-cu, Kidanemariam2013-xm, Uhlmann2014-ja}. In this study, we adopted the approach by~\citet{Kidanemariam2013-xm} where the fluid velocity in the vicinity of the particles is defined as the fluid in a spherical volume of diameter $2 d_p$ where the center of the spherical volume coincides with the particle center. Figure~\ref{fig:urms_Rep_fluid} shows the fluid velocity fluctuations as a function of $Re_p$. Overall, the trends are very similar to the particle velocity fluctuations in figure~\ref{fig:urms_Rep}. However, the magnitudes are smaller, indicating weaker fluid velocity fluctuations compared to the particles. It is likely that the inter-particle collisions leading to the particle velocity fluctuations occur over time scales that are too short for the fluid to respond, resulting in lower fluid velocity fluctuations.

\subsection{Autocorrelation and pairwise distributions} \label{sec:regimes}
\begin{figure}
\centerline{\includegraphics[width=\textwidth]{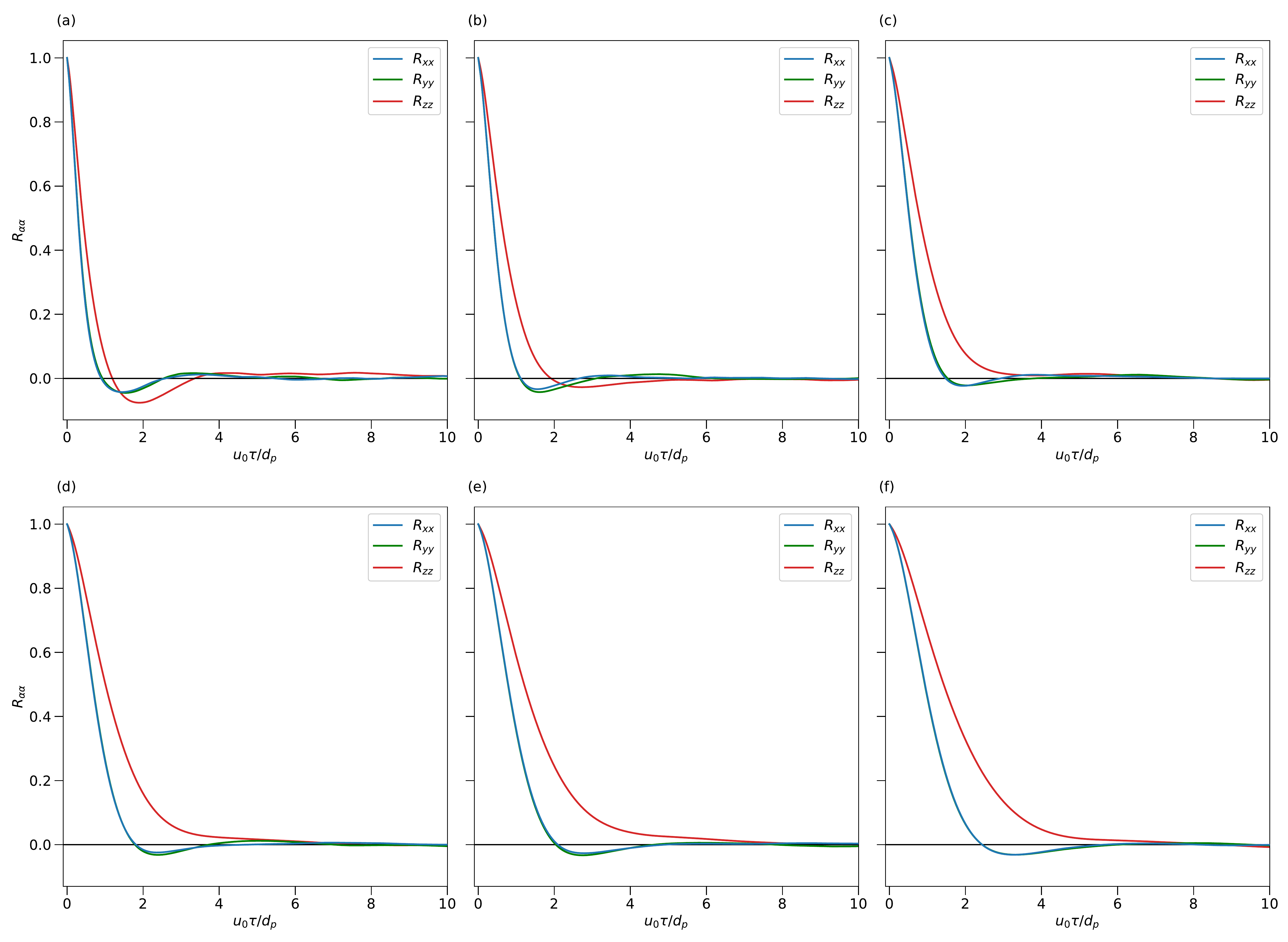}}%
\caption{\label{Rii_manu}Transverse ($R_{xx},R_{yy}$) and axial ($R_{zz}$)
 velocity fluctuation autocorrelation as a function of the lag time
 $\tau$ (equation~(\ref{all_Rii})) with different particle
 Reynolds numbers $Re_p$.(a) $Re_p=20$, (b) $30$, (c) $40$, (d)  $50$, (e) $60$, (f) $70$.}
\end{figure}

To understand the processes governing the fluctuating particle motions, we compute the instantaneous velocity fluctuation autocorrelation for a given time lag $\tau$

\begin{eqnarray}
\label{all_Rii} 
R_{\alpha\alpha}(\tau) = \dfrac{\ensmean{u^\prime_\alpha(t_0)u^\prime_\alpha(t_0 + \tau)}}{\ensmean{u^\prime_\alpha(t_0)u^\prime_\alpha(t_0)}},
\end{eqnarray}
and the results are shown in figure~\ref{Rii_manu}. For all $Re_p$, the transverse velocity fluctuations ($R_{xx},R_{yy}$) for $1 < u_0\tau/d_p < 6$ indicate regimes of anti-correlation, the extent of which decreases with increasing Reynolds number.
The axial velocity fluctuations $R_{zz}$ decorrelate monotonically to zero except for $Re_p = 20$ and $30$ where a region of anti-correlation exists. In addition, the transverse velocity fluctuations decorrelate faster than the axial velocity fluctuations because of preferential excitation of random particle motions in the axial direction by the axial flow. Similar results have been reported by previous three-dimensional
simulations and experimental results with different particle properties~\citep{Willen2019-rm, Esteghamatian2017-hc, Nicolai1995-wi}.
 
The decorrelation time for the $\alpha$ component of the velocity
fluctuations can be quantified by the true integral time scale
\begin{eqnarray}
\mathcal{T}_{\alpha,\infty} = \int_0^\infty R_{\alpha\alpha}(\tau) \ \dd \tau\,. \label{int_time_scale}
\end{eqnarray}
However, in simulations where data is limited, the computed integral time scale is instead approximated with
\begin{eqnarray}
\mathcal{T}_{\alpha,cal} = \int_0^{t_f} R_{\alpha\alpha}(\tau) \ \dd \tau\,, \label{eq:int_time_scale_approx}
\end{eqnarray}
where $t_f$ is the simulation time and $N_\tau = u_0 t_f/d_p$ is the nondimensional time to calculate the computed integral timescale. 
\begin{figure}
\centerline{\includegraphics[width=.45\textwidth]{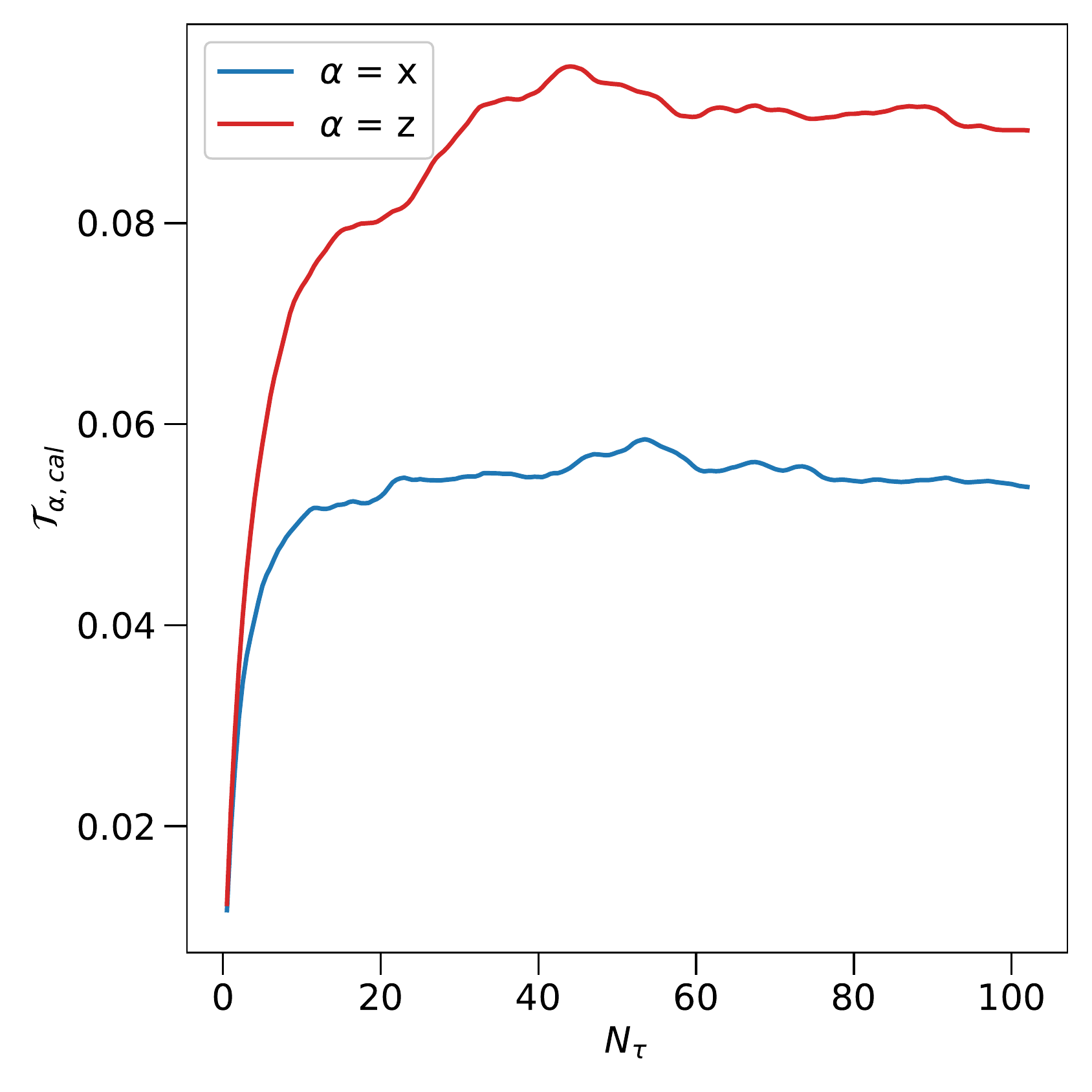}}%
\caption{\label{fig:T_Ntau} Computed integral timescale $\mathcal{T}_{\alpha,cal}$ as a function of $N_\tau$.}
\end{figure}
Figure~\ref{fig:T_Ntau} shows that the magnitude of $\mathcal{T}_{\alpha,cal}$ increases initially as a function of $N_\tau$ and fluctuates about a mean value due to the presence of wave-like autocorrelations. Similar wave-like autocorrelations were observed by~\citet{Esteghamatian2017-hc} and~\cite{Nicolai1995-wi}. To quantify the error associated with the computed integral timescale, we computed the mean and standard deviation of $\mathcal{T}_{\alpha,cal}$ as 
\begin{eqnarray}
&\mathbb{E}(\mathcal{T}_{\alpha,cal}) = \dfrac{1}{N_\tau-N_{\tau,\text{thresh}}} \sum_{i=N_{\tau,\text{thresh}}}^{N_\tau} \mathcal{T}_{\alpha,cal}^i, \\
&\sigma(\mathcal{T}_{\alpha,cal}) = \sqrt{ \mathbb{E}(\mathcal{T}^2_{\alpha,cal}) - \mathbb{E}(\mathcal{T}_{\alpha,cal})^2}, 
\end{eqnarray}
where $N_{\tau,\text{thresh}}$ is the minimum $N_{\tau}$ for $\mathcal{T}_{\alpha,cal}$ to reach equilibrium, and $\mathbb{E}(\mathcal{T}_{\alpha,cal})$ and $\sigma{(\mathcal{T}_{\alpha,cal})}$ are the mean and standard deviation of the calculated integral time scale, respectively. 

\begin{figure}
\centerline{\includegraphics[width=\textwidth]{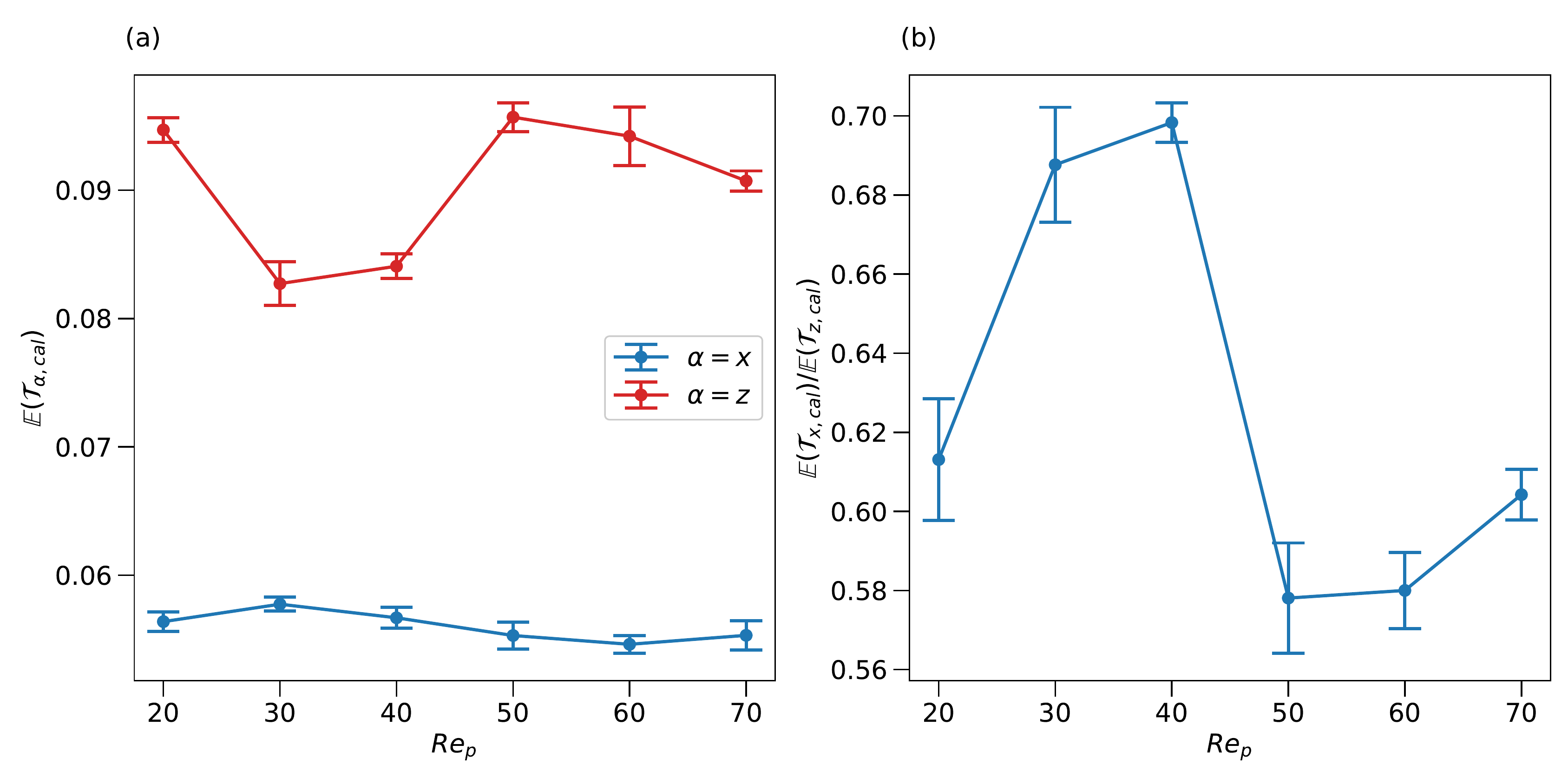}}%
\caption{\label{Timescale} 
 (a) The average integral time scales and (b) ratio of transverse integral timescale $\mathbb{E}(\mathcal{T}_{x,cal})$ to axial integral timescale $\mathbb{E}(\mathcal{T}_{z,cal})$ as a function of Reynolds number $Re_p$.}
\end{figure}

Figure~\ref{Timescale} shows the effect of $Re_p$ on $\mathbb{E}(\mathcal{T}_{\alpha,cal})$ and the ratio $\mathbb{E}(\mathcal{T}_{x,cal})/\mathbb{E}(\mathcal{T}_{z,cal})$, which is a measure of system anisotropy. Overall, $\mathbb{E}(\mathcal{T}_{z,cal}) > \mathbb{E}(\mathcal{T}_{x,cal})$ for the range of $Re_p$ simulated, implying $\mathbb{E}(\mathcal{T}_{x,cal})/\mathbb{E}(\mathcal{T}_{z,cal}) < 1$. This is consistent with the fact that the axial
velocity fluctuations take longer to decorrelate owing to the
presence of the mean flow~\citep{Esteghamatian2017-hc,Willen2019-rm,Nicolai1995-wi}. Interestingly, a minimum anisotropy (maximum of $\mathbb{E}(\mathcal{T}_{x,cal})/\mathbb{E}(\mathcal{T}_{z,cal})$) and maximum anisotropy (minimum of $\mathbb{E}(\mathcal{T}_{x,cal})/\mathbb{E}(\mathcal{T}_{z,cal})$) are observed at $Re_p = 40$ and $50$, indicating most and least efficient momentum transfer, respectively. In addition, a sharp decrease in $\mathbb{E}(\mathcal{T}_{x,cal})/\mathbb{E}(\mathcal{T}_{z,cal})$ is observed as $Re_p$ increases from $40$ to $50$. This is due to the presence of
competing mechanisms related to flow and particle physics, as discussed below.

\begin{figure}
\centerline{\includegraphics[width=\textwidth]{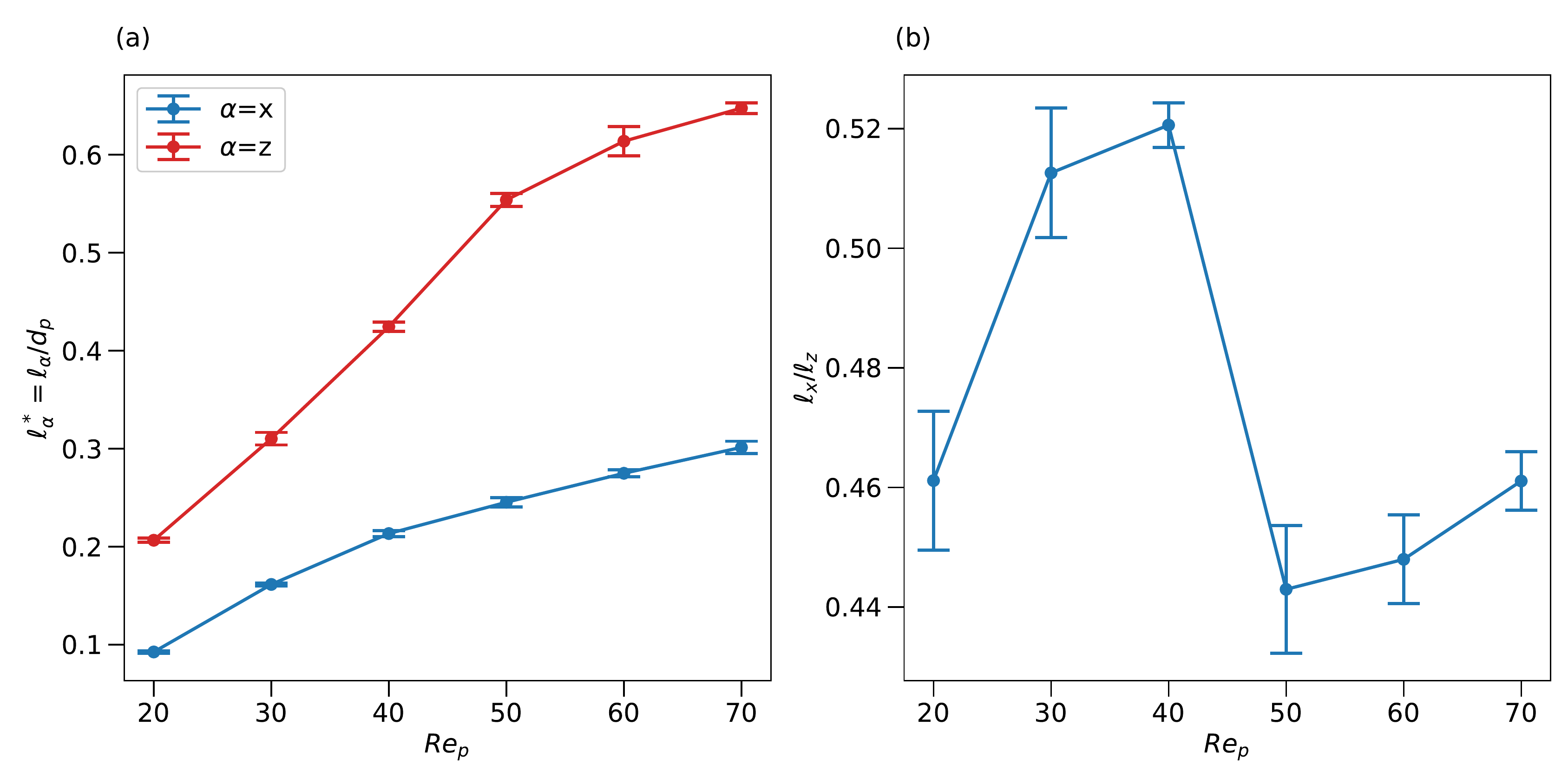}}
\caption{\label{fig:ell} Effect of particle Reynolds number on
 (a) the autocorrelation distance $\ell_\alpha^*=\ell_\alpha/d_p$ and (b) ratio of transverse to axial autocorrelation distance $\ell_x/\ell_z$.}
\end{figure}

We computed
the normalized autocorrelation length scale, $\ell_\alpha^* =
\ell_\alpha/d_p=u_{rms,\alpha}\mathbb{E}(\mathcal{T}_{\alpha,cal})/d_p$ which is a measure of the distance over which particle velocity fluctuations are still correlated, as shown
in figure~\ref{fig:ell}(a). When the particle velocity fluctuations are
dominated by collisions when the porosity is low, the effects of velocity
fluctuations do not propagate by more than the particle diameter, and thus $\ell_\alpha^* < 1$. However, when $\ell_\alpha^* > 1$ the particle velocity fluctuations are dominated by
the effects of the mean flow on each particle since the effects of velocity fluctuations propagate further
than a particle diameter. For all cases, both $\ell_x^*$ and $\ell_z^*$ are less than 1 indicating the particle velocity fluctuations are restricted by the relatively low porosity.~\citet{Nicolai1995-wi} reported $\ell_z^* > 1$ for particle suspension in the Stokes regime, while~\citet{Esteghamatian2017-hc} reported $\ell_z^* < 1$. We also computed $\ell_x/\ell_z$ which indicates anisotropy. As shown in figure~\ref{fig:ell}(b), the anisotropy is maximized at $Re_p=40$ and decreases sharply between $Re_p=40$ and $Re_p=50$, indicating an increase in the anisotropy. This agrees with figure~\ref{Timescale}(b) where a sharp decrease is also observed from $Re_p=40$ to $Re_p=50$. 

\begin{figure}
\centerline{\includegraphics[width=\textwidth]{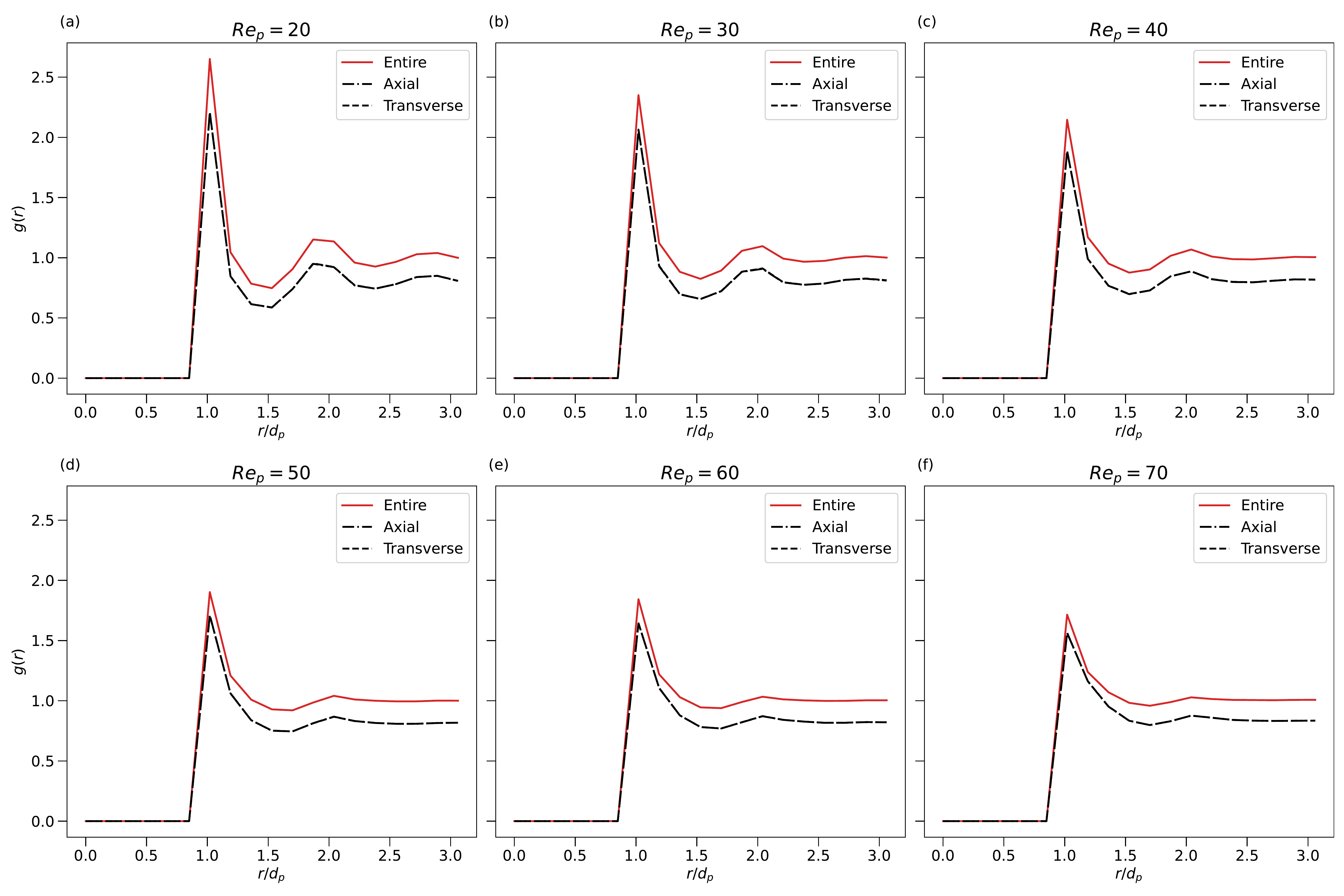}}%
\caption{\label{gr_example}The pairwise distribution function, $g(r,\theta)$ (equation~(\ref{g_r_theta})) averaged over three ranges for the simulated cases with different $Re_p$.}
\end{figure}

\begin{figure}
\centerline{\includegraphics[width=.45\textwidth]{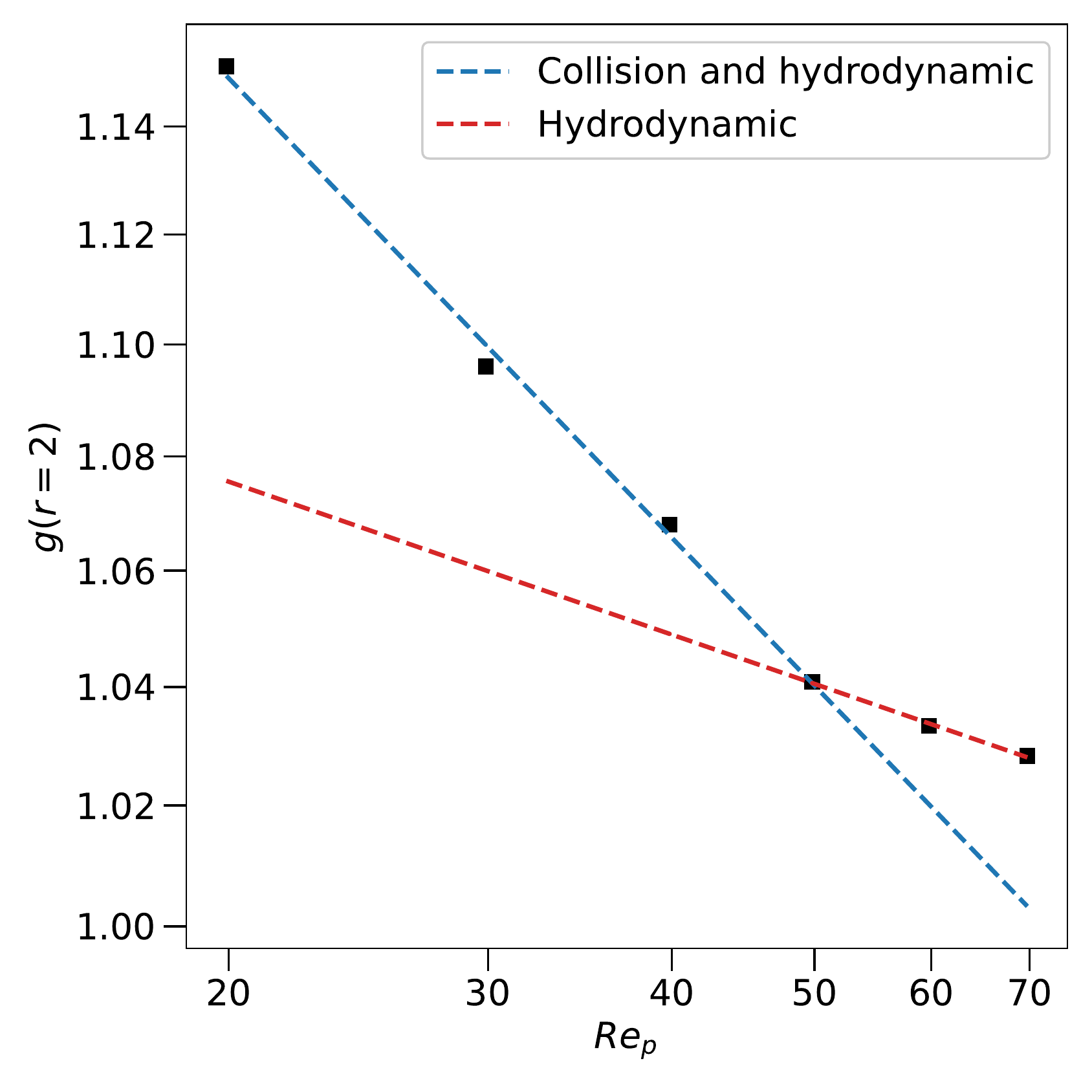}}%
\caption{\label{grRe} The radial distribution function at $r=2$, $g(r=2,\theta)$, averaged over $0\le \theta \le \pi/2$,  for ranges of $Re_p$ simulated, with different regimes indicated by the lines obtained through linear regression.}
\end{figure}

To highlight the different particle Reynolds number regimes leading to the anisotropy,
we adapt the pairwise-probability density function from~\citet{Yin2008-em}, which is given by
\begin{eqnarray}
g(r,\theta) = g(\boldsymbol{r}) = \frac{L_xL_y L_b}{d_p^3 N_p^2}\left< \sum_{i=1}^{N_p}\sum_{j=1,j\ne i}^{N_p} \delta(\boldsymbol{r} - \boldsymbol{r}_{ij})\right>, \label{g_r_theta}
\end{eqnarray}
where $L_b=z_t-z_b$ is the height of the bed where the porosity is homogeneous,
$\delta$ is defined as 
\begin{eqnarray}
\delta(\boldsymbol{r} - \boldsymbol{r}_{ij}) =  \label{delta_fun}
\begin{cases}
1 \ \ \ \ \boldsymbol{r} = \boldsymbol{r}_{ij}, \\
0\ \ \ \ \text{otherwise},
\end{cases}
\end{eqnarray} 
and $\boldsymbol{r}_{ij} = [r_{ij},
 \theta_{ij}]$ is the position vector between the center of particle $i$, $\boldsymbol{x}_i$, and
particle $j$, $\boldsymbol{x}_j$.  In cylindrical polar coordinates, 
$r_{ij} = \Vert\boldsymbol{x}_i - \boldsymbol{x}_j \Vert$ is the magnitude of
$\boldsymbol{r}_{ij}$ and 
\begin{eqnarray}
&& \theta_{ij} = \begin{cases}
\cos^{-1}[(\vert z_i -
 z_j\vert)/r_{ij}] & \text{if  } z_i -
 z_j < 0,\\ 
 \sin^{-1}[(\vert z_i -
 z_j\vert)/r_{ij}] & \text{if  } z_i -
 z_j \ge 0.
\end{cases}
\end{eqnarray}
We also compute the pairwise-probability distribution function or
radial distribution function as the angular average of $g(r,\theta)$
over three ranges: the entire range ($0 \le \theta \le \pi/2$), an effective axial range ($0 \le \theta
\le \pi/12$) and an  effective transverse range ($5\pi/12 \le \theta \le \pi/2$). As
shown in figure~\ref{gr_example}, the difference between
the transverse and axial radial distribution functions is a measure
of the particle arrangement preferences. The results indicate that particles do not have preferred arrangement over the $Re_p$ simulated, which is likely because the volume
fractions we simulate are too large~\citep{Willen2019-rm,Yin2007-eb}. At $r/d=1$ and $r/d=2$, there is an obvious increase in the magnitude of
the radial distribution function indicated by peaks in $g(r)$, an effect that decreases with
increasing particle Reynolds number (see figure~\ref{grRe}). This indicates that particles are
more likely to appear at these two locations. Comparing with the
three-dimensional simulations by~\citet{Willen2019-rm}, in which the peak
in the transverse $g(r)$ is slightly higher than the peak in the axial $g(r)$
(implying that a transverse arrangement of particle pairs is slightly favored
over an axial arrangement), we found no difference between the transverse and axial arrangements. A possible explanation of this observation is due the higher $Re_t$ simulated in which the effect of collisions is more prominent. As a result, momentum is transferred more efficiently from the axial to transverse directions, inducing higher transverse velocity fluctuations and disrupting the arrangement of transverse particle pairs observed by~\citet{Willen2019-rm}.

To further understand the effect of the particle
Reynolds number, figure~\ref{grRe} shows the effect of $Re_p$ on
$g(r/d_p = 2)$. For the cases simulated, two regimes can be identified
with regressions having correlation coefficients $R^2 > 0.99$.  In general, the
peak decreases with increasing Reynolds number, although the slope depends on two distinct flow regimes, which are referred to as Regime 1 ($Re_p < 50$) and Regime 2 ($Re_p \ge 50$) in what follows. 

\subsection{Particle-particle and fluid-particle interactions}

The mechanisms dictating the particle velocity fluctuations described in
Section~\ref{sec:regimes} can be explained through analysis
of the magnitude of stresses related to the particle-particle and particle-fluid interactions. 
The role of particle-particle collisions is to transfer
momentum from the axial direction to the transverse direction,
resulting in a more isotropic system~\citep{Esteghamatian2017-hc}. To quantify the magnitude of the stress induced by collisions and flow, we computed the normal contact stress $\sigma_{col,\alpha}$, normal lubrication stress $\sigma_{lub,\alpha}$, and hydrodynamic stress due to the fluid $\sigma_{hydro,\alpha}$ as a function of the vertical position in the fluidized bed. The lubrication stress is considered separately from collision and hydrodynamic stresses since both fluid-particle and particle-particle interaction are involved. A detailed derivation of these stresses can be found in Appendix~\ref{sec:appB}. All stresses are normalized by $d_p^2/\rho_f\nu^2$ and denoted as $\widehat{\sigma}$.

\begin{figure}
\centerline{\includegraphics[width=.45\textwidth]{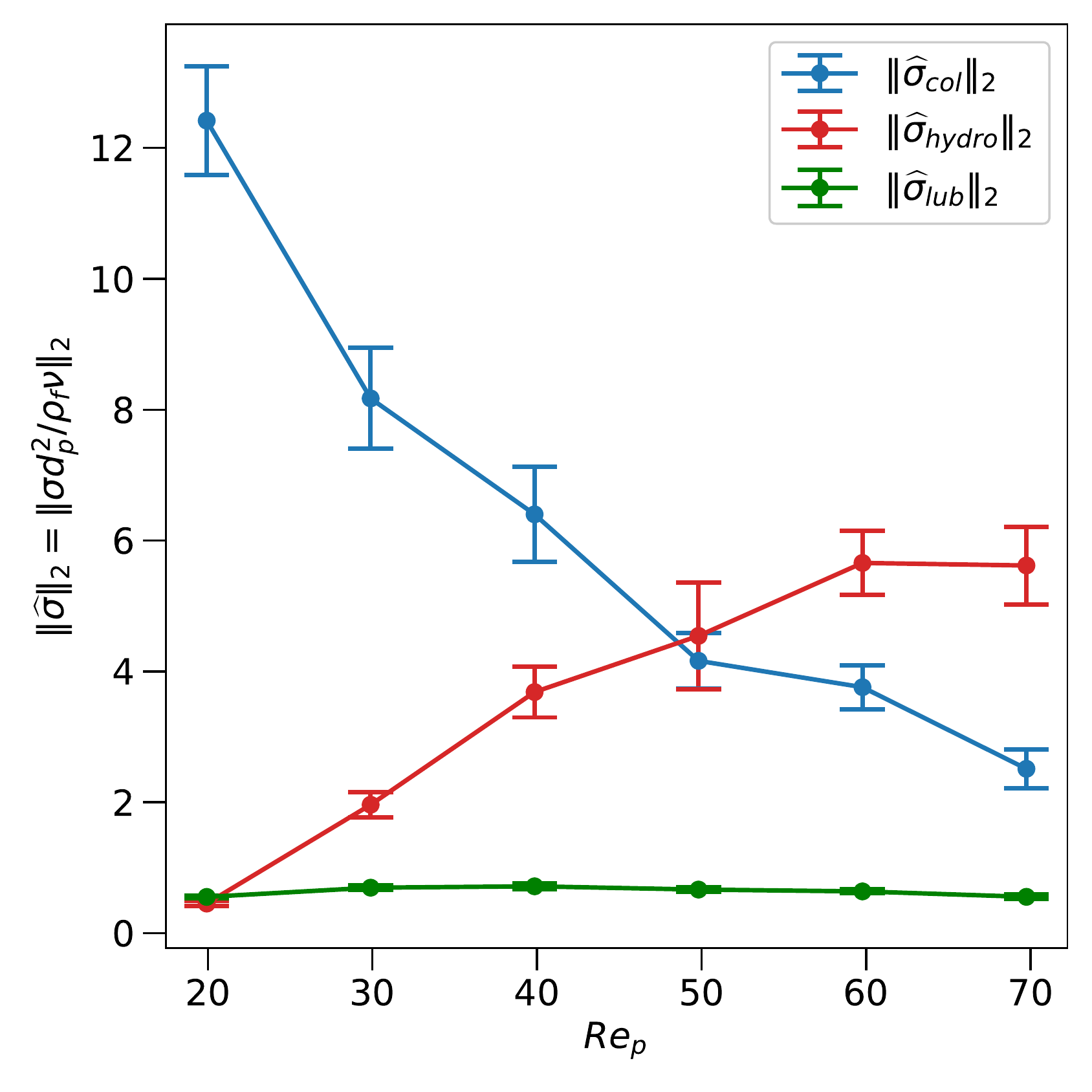}}%
\caption{\label{fig:Stress_Rep} The computed normal contact stress, normal lubrication stress and hydrodynamic stress as a function of particle Reynolds number $Re_p$.}
\end{figure} 
Overall, all the stresses fluctuate about a mean value away from the top and bottom boundaries of the fluidized bed. To quantify the effect of $Re_p$ on each component of the stress, we computed the vertical average of each stress first using the averaging operator defined in equation~\ref{mod_spa_average} and then computing the $\ell^2$ norm of the stress vector $\boldsymbol{\sigma}$ as $\norm{\boldsymbol{\sigma}}_2$. Figure~\ref{fig:Stress_Rep} shows the magnitude of the normalized mean stresses as a function of $Re_p$. As $Re_p$ increases, the effect of $\norm{\boldsymbol{\sigma}_{col}}_2$ decreases while the effect of $\norm{\boldsymbol{\sigma}_{hydro}}_2$ increases. Both~\citet{Zenit1997-wm} and~\citet{Derksen2007-mw} have found $\norm{\boldsymbol{\sigma}_{col}}_2$ decreases with decreasing $\phi$ beyond a critical volume fraction. The effect of $\norm{\boldsymbol{\sigma}_{lub}}_2$ is negligible, which also agrees with~\citet{Derksen2007-mw} who showed that the lubrication stress is negligible even though the simulated porosity and particle properties are different. 
\begin{figure}
\centerline{\includegraphics[width=.5\textwidth]{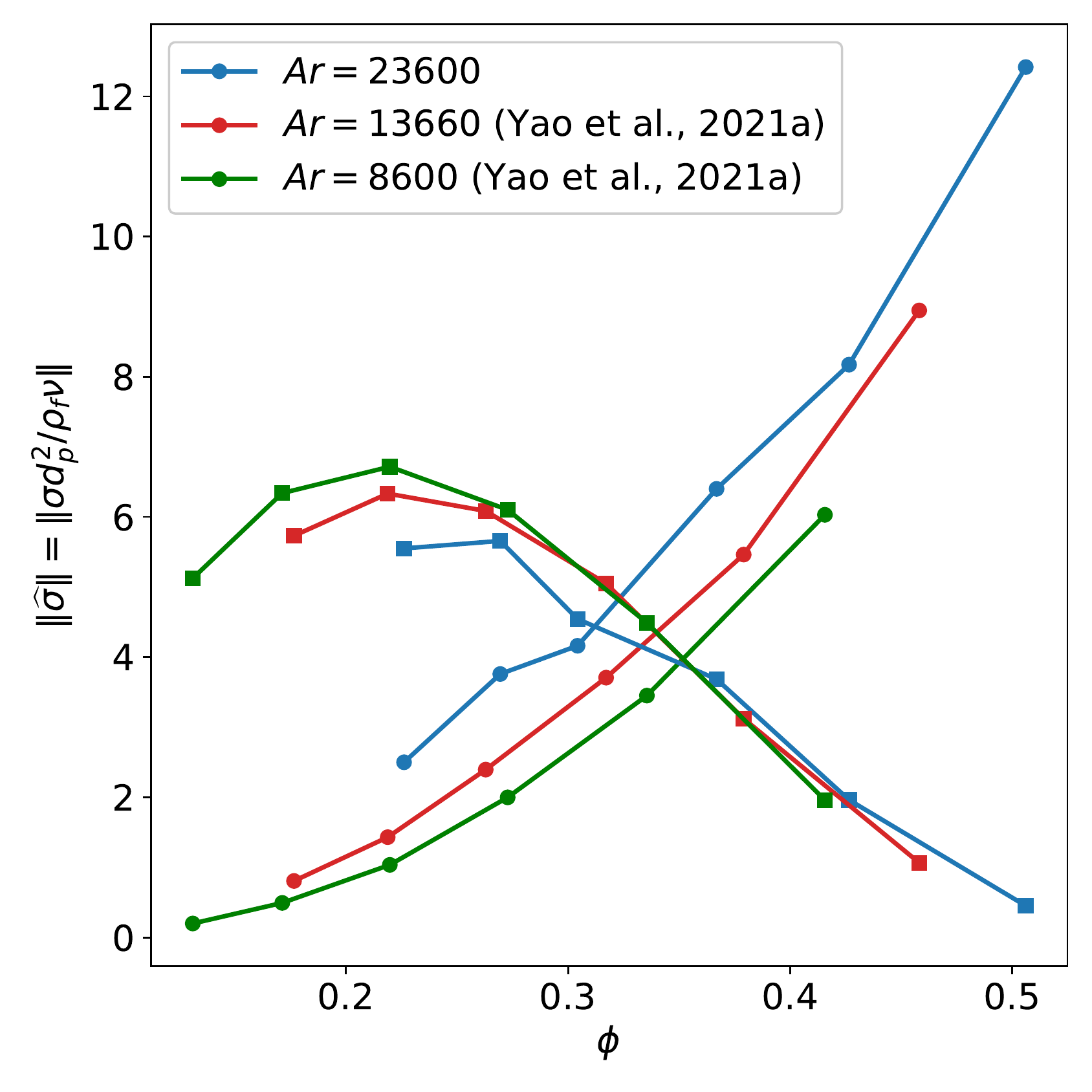}}%
\caption{\label{fig:stress_phi}The computed normal contact stress (circle) and hydrodynamic stress (square) as a function of volume fraction $\phi$ for Archimedes numbers of 8600, 13660 and 23600.}
\end{figure}

\begin{figure}
\centerline{\includegraphics[width=.9\textwidth]{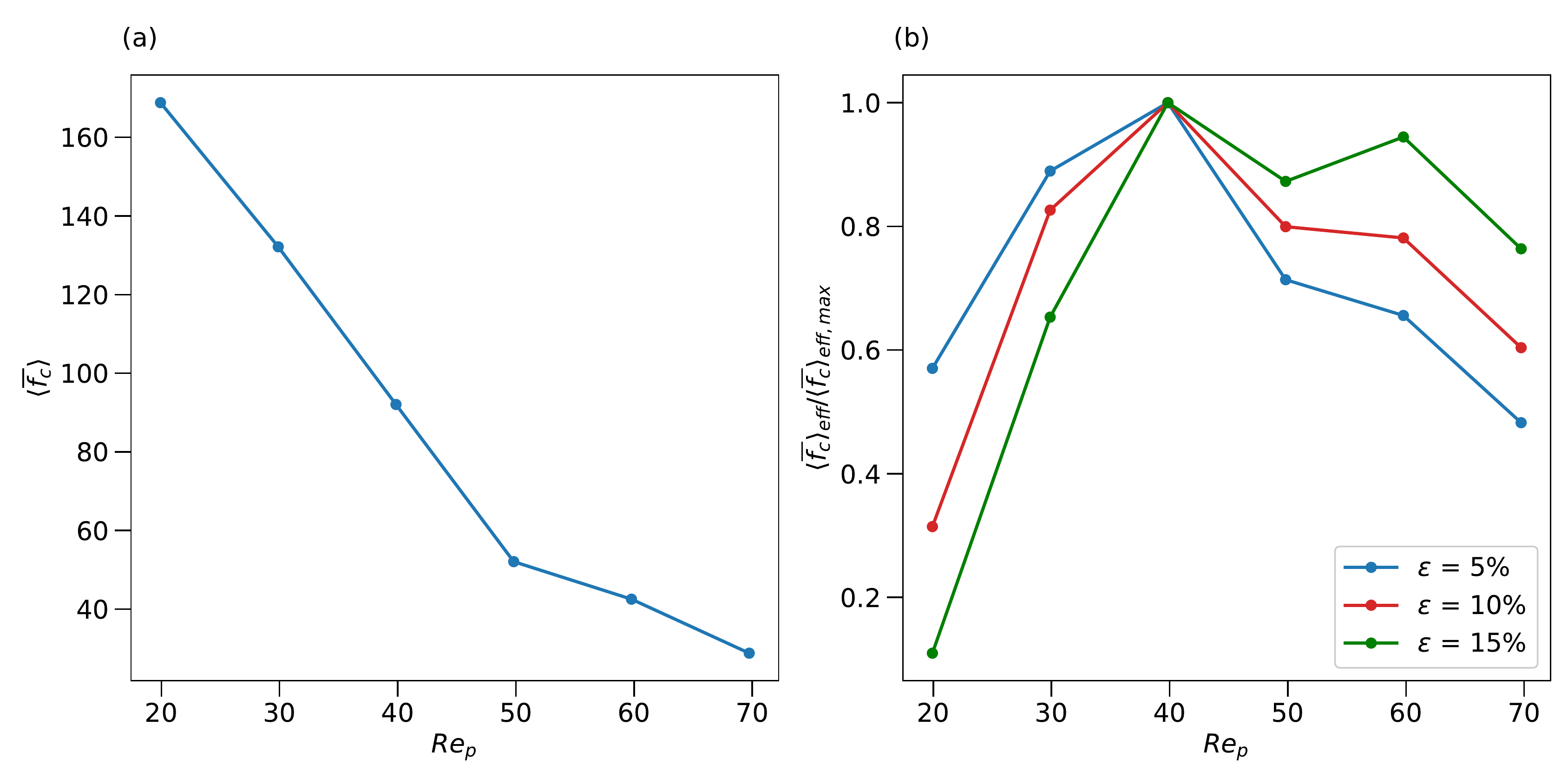}}%
\caption{\label{fig:col_freq}Time- and ensemble-averaged (a) collision
 frequency $\enstmean{f_c}$ and (b) effective collision frequency $\enstmean{f_c}_{\text{eff}}$ as a function of particle Reynolds number with different $\varepsilon$.}
\end{figure}

Comparing the magnitude of each stress (figure~\ref{fig:Stress_Rep}), at $Re_p = 20$ where $\norm{\boldsymbol{\sigma}_{col}}_2 >> \norm{\boldsymbol{\sigma}_{hydro}}_2$, collisions dominate over hydrodynamic effects and hydrodynamic effects are negligible. For $20 < Re_p < 40$, as $Re_p$ increases, hydrodynamic effects increase, $\norm{\boldsymbol{\sigma}_{col}}_2 > \norm{\boldsymbol{\sigma}_{hydro}}_2$ still persists but the difference is decreasing. In this regime, collisions still dominate over hydrodynamic effects but the relative importance of hydrodynamic effects increases. This corresponds to Regime 1 in figure~\ref{grRe} where both collisions and hydrodynamics are important in inducing velocity fluctuations. For $Re_p > 40$, $\norm{\boldsymbol{\sigma}_{hydro}}_2 > \norm{\boldsymbol{\sigma}_{col}}_2$ indicates hydrodynamic effects dominating over collisions, corresponding to Regime 2 identified in figure~\ref{grRe}. By comparing with results from~\citet{Yao2021-ky} for simulations of fluidized beds with Archimedes number of 8600 and 13660 (figure~\ref{fig:stress_phi}), we observe that the critical volume fraction (inversely proportional to particle Reynolds number) decreases as the Archimedes number increases. However, future work is required to establish a relationship between the optimum particle Reynolds number and a wider range of Archimedes numbers. 

The decreasing importance of the collisional stress with increasing $Re_p$ occurs because of a reduction in the collision frequency with increasing $Re_p$. We define a collision between two particles as occurring
when the separation distance between the particle centers
is less than the particle diameter $d_p$.
If the number of times particle $i$ collides with another particle during a simulation
time step $n$ is given by $N_{c,i}^n$, then the time- and
ensembled-average collision frequency is given by
\begin{eqnarray}
\enstmean{f_c} = \frac{1}{N_p(t_{max}-t_0)}\sum_{n=n_0}^{n_{max}}\sum_{i = 1}^{N_p} N_{c,i}^n\,,
\label{col_freq}
\end{eqnarray}
where $n_0=t_0/\Delta t$ and $n_{max}=t_{max}/\Delta t$. Since the
porosity and in turn the spacing between particles increases with
increasing $Re_p$ (figure~\ref{expan_index}), the likelihood of collisions should decrease with
increasing $Re_p$, leading to the monotonically-decreasing dependence of
$\enstmean{f_c}$ on $Re_p$ as shown in figure~\ref{fig:col_freq}(a). When the porosity is smaller, the likelihood of collision between particles is higher. 

For the simulated cases, the collision frequency as defined in equation~(\ref{col_freq}) is overestimated
for small $Re_p$ because particles may interact without colliding and producing a measurable collisional stress. To restrict collisions
to those with appreciable normal contact velocities, we define
the collision Stokes number with impact velocity $u_{imp}$ as
\begin{eqnarray}
\label{stokes_num1}
St_{imp} = \dfrac{u_{imp}\rho_p d_p}{9\rho_f \nu},
\end{eqnarray}
where $u_{imp}$ is the normal component of the relative particle velocities
contacting one another. The 
effective collision frequency, $\enstmean{f_c}_{\text{eff}}$, is
then computed by including collisions for which $St_{imp}>St_{thresh}$,
where $St_{thresh}$ is a threshold Stokes number. The threshold Stokes number is determined by setting a minimum rebound velocity with an empirical function relating the restitution coefficient to the Stokes number. Defining the velocity of a particle before it is subjected to lubrication and contact forces as $w_{t,\infty}$, the restitution coefficient is given by
\begin{eqnarray}
\label{eq:eps-St0}
&& \varepsilon = \dfrac{u_{rebound}}{u_{t,\infty}} = \varepsilon_{\max} \exp(-\dfrac{30}{St_{\infty}}).
\end{eqnarray}
Here, $\varepsilon_{\max} \approx 0.91 $ is the maximum restitution coefficient when $St_{\infty} \rightarrow \infty$~\citep{Legendre2006-tb}, where 
\begin{eqnarray}
&& St_{\infty} = \dfrac{u_{t,\infty}d_p \rho_p}{9\rho_f \nu_f}.
\end{eqnarray}
In this work, since $u_{t,\infty}$ is difficult to quantify due to simultaneous collisions with the soft-sphere modeling approach, we rearrange equation~\ref{eq:eps-St0} to relate $St_{imp}$ to $\varepsilon$ by using the dry restitution coefficient $e_{dry} = u_{rebound}/u_{imp}$ such that 
\begin{eqnarray}
&& St_{imp} = \dfrac{30\varepsilon}{e_{dry} (\log \varepsilon - \log \varepsilon_{\max})} = \dfrac{u_{imp}d_p \rho_p}{9\rho_f \nu_f}.
\end{eqnarray}
To understand the effect of different $\varepsilon$ on $St_{thresh}$, figure~\ref{col_freq}(b) shows the normalized effective collision frequency as a function of $Re_p$ for different $\varepsilon$. In general, as $\varepsilon$ increases, the effective collision frequency decreases and the maximum effective collision frequency shifts to higher $Re_p$. However, for the range of  $\varepsilon$ tested, the maximum effective collision frequency occurs when $Re_p = 40$, indicating that the maximum effective collision frequency is not very sensitive to $\varepsilon$. Furthermore, the maximum effective collision frequency
coincides with the minimum anisotropy observed in
figure~\ref{Timescale} and figure~\ref{fig:ell}, indicating a point in which momentum transfer
from the axial to transverse directions is optimal. 

\section{Summary and conclusions}
We studied the effects of the particle Reynolds number on the behavior of monodispersed spherical particles in PRS simulations of a three-dimensional, liquid-solid fluidized bed. The particle Reynolds number was varied by varying the flow rate suspending particles in the axial ($z$) direction. Analysis of various statistics provided insights into the observed particle motions. Boundary effects associated with the fluidized bed are identified and excluded in statistical calculations to improve accuracy. The wave modes are studied in both physical and spectral spaces to shed light on the source of the particle velocity fluctuations. By separating volume fraction fluctuations into low and high wavenumber components, wavelike behavior is clearly observed. For low particle Reynolds number ($Re_p = 20$), waves with both low and high wavenumbers are strongly dependent on vertical position and weakly dependent on time. As particle Reynolds number increases, waves with low wavenumber depend strongly both on vertical positions and time while wave motions with high wavenumber are less apparent. Volume fraction fluctuations in spectral space further reveal that high-frequency wave modes are more significant at low particle Reynolds number. Low wavenumber modes can be estimated well with the kinematic wave relationship using linear regression while no discernible kinematic relationship can be observed for high wavenumber modes, indicating that the high wavenumber fluctuations are probably random. Analysis of the root-mean-square particle velocity fluctuations indicates a maximum at an intermediate particle Reynolds number ($Re_p = 40$) in both the transverse and axial directions.

The decorrelation timescales in the axial and transverse directions reveal that momentum transfer from the axial to the transverse directions is most efficient at $Re_p=40$ and least efficient at $Re_p=50$. A sharp decrease in the efficiency of the momentum transfer is observed as $Re_p$ increases from 40 to 50, revealing a transition in the flow regime. Because the length scale over which particle motions decorrelate was less than $1 d_p$ for all particle Reynolds numbers simulated. the transition is dominated by porosity effects. By analyzing the pairwise distribution function, we found that the probability that particle pairs are aligned at a distance of $2d_p$ decreases with increasing particle Reynolds number. The rate of this decrease revealed two distinct regimes that are consistent with the momentum transfer regimes discussed above. 

To understand the mechanisms controlling the flow regimes and the sharp decrease in momentum transfer from the axial to transverse directions, we computed average collision and hydrodynamic stresses as a function of $Re_p$. The results indicate that the relative magnitude of collision to hydrodynamic effects controls the efficiency of inducing particle velocity fluctuations, momentum transfer and particle alignment. For $20 < Re_p \le 40$, collisions dominate over hydrodynamic effects but the relative importance of hydrodynamic effects increases, indicating a co-existence of mechanisms related to flow and collisions (Regime 1) that leads to the peak in particle velocity fluctuations and momentum transfer. As the particle Reynolds number increases ($Re_p > 40$), hydrodynamic effects dominate over collision effects (Regime 2). Due to a lack of effective collisions, particle velocity fluctuations decrease and a sharp decrease in momentum transfer efficiency is observed. The lack of collisions arises from a decrease in the collision frequency with increasing $Re_p$. We found that it was important to quantify the collision frequency by an effective collision frequency based on collisions satisfying a threshold Stokes number. Defined this way, the effective collision frequency peaks at an $Re_p$ that coincides with that of the highest particle fluctuations and a sharp decrease in momentum transfer. 

Our results imply biofilm detachment models in fluidized-bed reactors should focus on collision effects for $Re_p \le Re_{p,\alpha}$, collision and hydrodynamic effects for $Re_{p,\alpha} < Re_p < Re_{p,\beta}$, and hydrodynamic effects for $Re_p \ge Re_{p,\beta}$ where $Re_{p,\alpha} \approx 40$ and $Re_{p,\beta} \approx 50$ for an Archimedes number of 23600. This study excludes the effect of adhesive force on biofilm and Archimedes number. Further work is required to quantify the effect of adhesive force on particle dynamics and establish a relationship between the optimum particle Reynolds number and Archimedes number. Furthermore, our results imply that mixing within liquid-solid fluidized bed reactors is likely to be optimized at an intermediate $Re_p$ at which particle velocity fluctuations are expected to be the strongest. Indeed, previous fluidized bed reactor studies with a large Archimedes number show that treatment performance is optimized at a similar intermediate $Re_p\approx 30-40$~\citep{Jaafari2014-ay}. We anticipate that the results of this study will inform fluidized-bed reactor design and modeling for domestic and industrial wastewater treatment, enabling more reliable and energy-efficient operation.
\\ \\
\textbf{Acknowledgments.} This work used the Extreme Science and Engineering Discovery Environment (XSEDE), which is supported by National Science Foundation grant number ACI-1548562. Simulations were conducted with supercomputer resources under XSEDE Project CTS190063. The authors acknowledge the Texas Advanced Computing Center (TACC) at The University of Texas at Austin for providing HPC resources that have contributed to the research results reported within this paper. We thank Hyungoo Lee and Sivaramakrishnan Balachandar
from the University of Florida for providing us with their IBM code. We also thank Edward Biegert, Bernhard Vowinckel, Thomas Köllner and Eckart Meiburg from the University of California, Santa Babara, for assistance with implementation of the collision models.
\\ \\
\textbf{Funding.} This work was funded by the California Energy Commission (CEC) under CEC project number EPC-16-017, the U.S. NSF Engineering Center for Reinventing of the Nation’s Urban Water Infrastructure (ReNUWIt) under Award No. 1028968, and Office of Naval Research Grant N00014-16-1-2256. This document was prepared as a result of work sponsored in part by the California Energy Commission. It does not necessarily represent the views of the Energy Commission, its employees, or the State of California. Neither the Commission, the State of California, nor the Commission’s employees, contractors, or subcontractors makes any warranty, express or implied, or assumes any legal liability for the information in this document; nor does any party represent that the use of this information will not infringe upon privately owned rights. This document has not been approved or disapproved by the Commission, nor has the Commission passed upon the accuracy of the information in this document.
\\ \\
\textbf{Declaration of interests.} The authors report no conflict of interest.
\\ \\
\textbf{Author ORCID.}\\
Yinuo Yao https://orcid.org/0000-0001-8328-6072 \\
Craig Criddle https://orcid.org/0000-0002-2750-8547 \\
Oliver Fringer https://orcid.org/0000-0003-3176-6925

\appendix
\section{Instantaneous Eulerian volume fraction}
\label{sec:appA}
In our simulations, the instantaneous Eulerian volume fraction $\phi(\boldsymbol{x},t)$ can be estimated using a second-order level-set approximation~\citep{Kempe2012-lp}. Defining the particle center as $\boldsymbol{x}_p$ and Eulerian grid as $\boldsymbol{x}_{ijk}$ where $i$, $j$ and $k$ represent each direction. The volume fraction of each grid cell is defined as 
\begin{eqnarray}
\phi(x_i, y_j, z_k) = \dfrac{\sum_{l=1}^{8} -\psi_m \mathcal{H}(-\psi_m)}{\sum_{l=1}^{8} \abs{\psi_m}},
\end{eqnarray}
where $l$ is an integer representing the corner of an Eulerian grid, $\psi$ is the level-set function for spheres that is defined as 
\begin{eqnarray}
 && \psi(\boldsymbol{x}, \boldsymbol{x}_p, r_p) = \dfrac{\Vert \boldsymbol{x} - \boldsymbol{x}_p \Vert_2}{r_p} - 1
\end{eqnarray}
and $\mathcal{H}(-\psi_m)$ is the Heaviside function 
\begin{eqnarray}
 \mathcal{H}(\alpha) = \begin{cases}
 0, & \alpha \le 0, \\
 1, & \alpha > 0.
 \end{cases} 
\end{eqnarray}
To ensure the accuracy of level-approximation for $\phi(\boldsymbol{x},t)$, we use a grid spacing equivalent to the simulations where $h = d_p/25.6$.

\section{Determination of particle-related stresses}
\label{sec:appB}
The governing equation of particle motion can be described as 
\begin{eqnarray}
\label{eq:total_force}
  m_p \dv{\boldsymbol{u}_p}{t} = \boldsymbol{F}_{h,p} + \boldsymbol{F}_{col,p} + \boldsymbol{F}_{lub,p} - V_p(\rho_p - \rho_f)\boldsymbol{g} , 
\end{eqnarray}
where $\boldsymbol{u}_p$ is the translational velocity of a particle, $\boldsymbol{F}_{h,p}$ is the drag force on the particle, $\boldsymbol{F}_{col,p}$ and $\boldsymbol{F}_{lub,p}$ are, respectively, the collision and normal lubrication force on particle $p$. Fluidization occurs when the weight of particle is balanced by the average drag force. Therefore, the drag force $F_{h,p}$ can be decomposed into two components such that
\begin{eqnarray}
\label{eq:drag_decomp}
 \boldsymbol{F}_{h,p} = \overline{\boldsymbol{F}_{h,p}} + \boldsymbol{F}_{h,p}^\prime = V_p(\rho_p - \rho_f)\boldsymbol{g} + \boldsymbol{F}_{h,p}^\prime,
\end{eqnarray}
where $\overline{\boldsymbol{F}_{h,p}}$ is the drag force that balances the weight of particle and $\boldsymbol{F}_{h,p}^\prime$ is the fluctuation force that results in acceleration. By substituting equation~\ref{eq:drag_decomp} into equation~\ref{eq:total_force}, the governing equation can be simplified as 
\begin{eqnarray}
   m_p \dv{\boldsymbol{u}_p}{t} = \boldsymbol{F}_{h,p}^\prime + \boldsymbol{F}_{col,p} + \boldsymbol{F}_{lub,p}.
\end{eqnarray}
In this study, the stresses due to $ \boldsymbol{F}_{h,p}^\prime$, $\boldsymbol{F}_{lub,p}$ and $\boldsymbol{F}_{col,p} $  are defined as hydrodynamic, lubrication and collision stresses respectively. Since lubrication stress arises when two or more particles move closer to one another due to particle-particle interaction, therefore, lubrication stress is considered separately from the hydrodynamic stress in which the particle motion is only affected by the interaction between fluid and the particle.

The normal lubrication force and collision forces are only nonzero when the separation distance $\zeta_n$ between particle centers are less than $d_p+2h$ and $d_p$ respectively. As such, the governing equation of particle motion can be rewritten as
\begin{eqnarray}
\label{eq:total_force_mod}
  m_p \dv{\boldsymbol{u}_p}{t} = 
  \begin{cases}
   \boldsymbol{F}_{h,p}^\prime, & \text{ for }  \Vert \boldsymbol{x}_p - \boldsymbol{x}_q \Vert_2 > d_p + 2h, \\
  \boldsymbol{F}_{h,p}^\prime + \boldsymbol{F}_{lub,p}, & \text{ for }  d_p <  \Vert \boldsymbol{x}_p - \boldsymbol{x}_q \Vert_2 \le d_p + 2h,\\
  \boldsymbol{F}_{col,p}, & \text{ for } \Vert \boldsymbol{x}_p - \boldsymbol{x}_q \Vert_2 \le d_p,
  \end{cases}
\end{eqnarray}
where $x_p$ and $x_q$ are the particle center positions. 

To determine the stresses as a function of vertical position in the domain, we first discretize the domain into slices with a vertical spacing of $d_p/2$. If we assume the acceleration of a particle at time step $n+1/2$ is given by 
\begin{eqnarray}
 \boldsymbol{a}_p^{n+1/2} = \dfrac{\boldsymbol{u}_p^{n+1} -  \boldsymbol{u}_p^{n}}{\Delta{t}}.
\end{eqnarray}
Then the total force $\boldsymbol{F}_{t,p}$ at time step $n+1/2$ is $m_p\boldsymbol{a}_p^{n+1/2}$. To demonstrate the validity of equation~\ref{eq:total_force_mod}, we define an average error metric 
\begin{eqnarray}
 \ensmean{\boldsymbol{\varepsilon}_{f}} = \dfrac{1}{N_p }\sum_{i=1}^{N_p} \dfrac{m_p\boldsymbol{a}_i^{n+1/2}}{\boldsymbol{F}^n_{i,rhs}} - 1,
\end{eqnarray}
where $\boldsymbol{F}^n_{i,rhs}$ is the summation of all forces experienced by the particle $i$ at time $n$. As shown in table~\ref{tab:stress_closure}, the mean error associated with the approximation is less than 1\% for all cases, demonstrating the validity of equation~\ref{eq:total_force_mod}.

\begin{table}
\begin{center}
\begin{tabular}{ccccccc}
\hline 
$Re_p$ & 20 & 30 & 40 & 50 & 60 & 70 \\  \hline
$\enstmean{\varepsilon_{f}}_x \times 10^{2} $ & 0.06 $\pm$ 2.17 & 0.14 $\pm$ 1.88 & 0.16 $\pm$ 1.76 & 0.21 $\pm$ 2.80 & 0.27 $\pm$ 1.97 & 0.26 $\pm$ 2.01 \\
$\enstmean{\varepsilon_{f}}_y \times 10^{2} $ & 0.06 $\pm$ 2.14 & 0.12 $\pm$ 1.91 & 0.17 $\pm$ 1.76 & 0.21 $\pm$ 2.83 & 0.26 $\pm$ 1.93 & 0.25 $\pm$ 1.93 \\
$\enstmean{\varepsilon_{f}}_z \times 10^{2} $ & 0.05 $\pm$ 2.17 & 0.12 $\pm$ 1.95 & 0.17 $\pm$ 1.78 & 0.24 $\pm$ 2.95 & 0.28 $\pm$ 2.02 & 0.26 $\pm$ 2.07 \\ \hline
\end{tabular}%
\caption{
  \label{tab:stress_closure} Error $\boldsymbol{\varepsilon_{f}}$ associated with force balance in equation~\ref{eq:total_force_mod}.}
  \end{center}
\end{table}

The hydrodynamic force at time step $n+1/2$ can be determined as 
\begin{eqnarray}
 \boldsymbol{F}_{h,p}^\prime = 
 \begin{cases}
 \boldsymbol{F}_{t,p}, & \text{ for } \Vert \boldsymbol{x}_p - \boldsymbol{x}_q \Vert_2 > d_p + 2h,\\
 \boldsymbol{F}_{t,p}- \boldsymbol{F}_{lub,p} , & \text{ for } d_p <  \Vert \boldsymbol{x}_p - \boldsymbol{x}_q \Vert_2 \le d_p + 2h,\\
 \end{cases}
\end{eqnarray}
and the hydrodynamic stress over particle surface area $(\sigma_{hydro,p})^{n+1/2} = \boldsymbol{F}_{h,p}^\prime/(\pi d_p^2)$ can be calculated. The hydrodynamic stress for a particle bin $k$ can be calculated as
\begin{eqnarray}
 (\boldsymbol{\sigma}_{hydro})_k =\dfrac{1}{N} \sum_{n=1}^{N_p} \sum_{t=1}^{N_t} \dfrac{(\vert \boldsymbol{F}_{h}^\prime\vert)^{t}_i}{\pi d_p^2}\mathbf{1}_{z_{l} < z_i < z_{u}}(z_i) ,
\end{eqnarray}
where $N = \sum^{N_t N_p}\mathbf{1}_{z_{l} < z_i < z_{u}}(z_i) $ is the number of samples in each bin, $z_i$ is the vertical position of particle $i$, $\boldsymbol{e}$ is vector of ones and $\mathbf{1}_{z_{l} < z_i < z_{u}}(z_i)$ is the indicator function that is defined as
\begin{eqnarray}
\mathbf{1}_{z_{l} < z_p < z_{u}}(z_i) = 
\begin{cases}
1 & z_l < z_p < z_u, \\
0 & \text{otherwise.}
\end{cases}
\end{eqnarray}

In the simulations, the collision model is based on the Adaptive Collision Time Model (ACTM) proposed by~\citet{Kempe2012-lp} and tangential collision models by~\cite{Biegert2017-ku}. In ACTM, each collision is assumed to occur over $10\Delta{t}$ instead of $\Delta{t}$ in the soft-sphere collision model. At each time step, the normal contact force $\boldsymbol{F}_{col}$ is determined and the contributions of all collisions at a bin $k$ is determined with 
\begin{eqnarray}
 (\boldsymbol{\sigma}_{col})_k = \dfrac{1}{N} \sum_{n=1}^{N_p} \sum_{t=1}^{N_t} \dfrac{(\vert \boldsymbol{F}_{con} \vert)^{t}_i}{\pi d_p^2}\mathbf{1}_{z_{l} < z_i < z_{u}}(z_i).
\end{eqnarray}

A similar procedure can be applied to the lubrication stress after assuming the lubrication forces are also stretched over several time steps. The contributions of all lubrication forces in a bin $k$ is determined with 
\begin{eqnarray}
 (\boldsymbol{\sigma}_{lub})_k = \dfrac{1}{N} \sum_{n=1}^{N_p} \sum_{t=1}^{N_t} \dfrac{(\vert \boldsymbol{F}_{lub}\vert)^{t}_i}{\pi d_p^2}\mathbf{1}_{z_{l} < z_i < z_{u}}(z_i) .
\end{eqnarray}

\end{document}